# A Two-Stage Bayesian Framework for Multi-Fidelity Sequential Updating of Spatial Fragility Fields

Abdullah M. Braik[a] and Maria Koliou[b*]

[a)] Assistant Professor, Department of Civil and Environmental Engineering, Old Dominion University, Norfolk, VA 23529, U.S.A.; formerly Postdoctoral Research Associate, Zachry Department of Civil and Environmental Engineering, Texas A&M University, College Station, TX 77843, U.S.A., E-mail: abraik3@tamu.edu

[b)] Associate Professor and Zachry Career Development Professor II, Zachry Department of Civil and Environmental Engineering, Texas A&M University, College Station, TX, 77843, U.S.A., E-mail: maria.koliou@tamu.edu (*Corresponding author)

# A Two-Stage Bayesian Framework for Multi-Fidelity Sequential Updating of Spatial Fragility Fields

**Abstract**

This paper develops a Bayesian framework for updating event-conditioned spatial fragility fields by combining physics-informed priors with incrementally available post-disaster observations. The framework supports sequential refinement of regional vulnerability estimates as additional post-disaster evidence becomes available. The framework reformulates physics-informed fragility estimates into a Probit–Normal (PN) representation that captures aleatory variability and epistemic uncertainty in an analytically tractable form. Stage 1 performs local Bayesian updating by moment-matching PN marginals to Beta surrogates that preserve their probability shapes, enabling conjugate Beta–Bernoulli updates with soft, multi-fidelity observations. Fidelity weights encode source reliability, and the resulting Beta posteriors are re-projected into PN form, producing heteroscedastic fragility estimates whose variances reflect data quality and coverage. Stage 2 assimilates the heteroscedastic pseudo-observations through an ordered probit-warped Gaussian Process (GP) that propagates information across space and structural archetypes. A shared latent spatial severity field and strictly positive threshold-gap functions enforce pathwise ordering of exceedance probabilities and yield nonnegative exact damage-state probabilities. The framework is applied to the 2011 Joplin tornado, where wind-field priors and computer-vision damage assessments are fused under varying assumptions about tornado width, sampling strategy, and observation completeness. Controlled synthetic experiments and the Joplin application show that the framework can reduce the influence of biased priors as evidence accumulates, propagate locally updated information into unobserved regions, and produce uncertainty-aware, ordinally valid exceedance and exact damage-state probability fields.

# 1. Introduction

## *1.1. Motivation and Problem Setting*

Natural hazards continue to challenge communities worldwide, revealing persistent gaps in our ability to anticipate, characterize, and interpret structural damage. Despite advances in engineering design, hazard modeling, and resilience planning, the observed performance of the built environment under extreme loading frequently diverges from model predictions. These discrepancies stem from substantial aleatory uncertainties in hazard intensity and structural capacity, as well as epistemic uncertainties embedded in the assumptions used to represent them.

Before hazard, risk assessments rely on probabilistic vulnerability models derived from physics-based simulations, empirical data from past events, heuristic rules, or hybrid formulations. These models are indispensable for mitigation and preparedness planning, yet their usefulness is largely confined to the pre-event phase. Once a hazard unfolds, the pre-event regional damage prediction field remains conditional on the hazard characterization, fragility assumptions, and inventory information available before the event. As event-specific observations become available, this prior field can be sequentially refined to reflect the realized damage pattern and its remaining uncertainty.

After a hazard, actual damage has occurred, and pre-event regional damage predictions may become insufficient once event-specific observations begin to emerge. Field inspections, post-disaster imagery, aerial surveys, and crowdsourced reports provide direct evidence of building performance across affected regions. However, these data streams are often incomplete, delayed, and heterogeneous in reliability. A remaining challenge is to assimilate such evidence sequentially into prior regional predictions while representing uncertainty and propagating information to locations without direct observations.

As a result, pre-event vulnerability models and post-event observations have progressed along parallel tracks, even though the limitations of one align almost perfectly with the strengths of the other. Without a unified framework to connect them, vulnerability models remain disconnected from the evidence that could refine them, while observational data lacks a coherent prior for rigorous inference. This disconnect represents a missed opportunity to develop dynamic digital twins that operate before and after a hazard, adapt to varying data fidelity and coverage, and provide a principled mathematical foundation for sequential decision support.

### *1.2. Background on the Development of Fragility Models Within Probabilistic Risk Assessment*

Fragility modeling emerged from the evolution of structural reliability theory, which reframed structural safety as a probabilistic event influenced by uncertainty in both loading and resistance. Early foundational work introduced limit-state formulations, reliability indices, and probability-of-failure concepts, enabling structural performance to be quantified probabilistically rather than through fixed safety factors [1-3]. These ideas permeated modern design practice through load- and resistance-factor methodologies [4], shifting vulnerability assessment away from deterministic margins toward explicit representations of uncertainty. Within this context, fragility functions emerged as practical tools for expressing the probability that hazard demand exceeds structural capacity [5], accelerating adoption during the rise of performance-based earthquake engineering. Early seismic applications, commonly modeled using lognormal exceedance curves, supported scalable regional loss estimation [6, 7] and informed methodologies that remain central to contemporary practice [8].

Over the past two decades, fragility modeling has expanded across hazards, building types, and infrastructure systems [9]. In earthquake engineering, empirical, experimental, and simulation-based methods have produced multi-state fragilities for reinforced concrete systems [10-13], steel systems [14-16], wood systems [17, 18], and nonstructural components [19]. Incremental dynamic analysis [20] further strengthened the statistical treatment of record-to-record variability and improved the robustness of fragility modeling.

Fragility concepts were subsequently extended to other hazards. For hurricanes, fragility functions have been developed for wind effects on low-rise residential structures [21-23], flood impacts [24-26], and storm-surge or wave-induced damage [27, 28]. Tornado-specific fragilities have been generated for buildings subjected to highly localized extreme wind fields [29-31]. Beyond buildings, fragility modeling has informed risk assessments for bridges [32-34], electric power networks [35-38], and pipeline systems [39, 40].

Today, fragility functions support a wide range of applications, including regional loss estimation [41], performance-based engineering [42], and mitigation planning at the community scale [43]. Yet even the most advanced fragility models remain fundamentally pre-event constructs. Their mathematically structured form positions them naturally as priors for post-event inference, but this capability remains largely unrealized. Progress in this direction requires frameworks that treat fragility functions as updatable Bayesian priors, capable of continuous refinement as heterogeneous, multi-fidelity observations become available after a hazard.

### *1.3. Background on Post-Disaster Data and Damage Assessment*

Rapid and reliable post-disaster damage assessment is essential for emergency response, loss estimation, and early recovery planning. Traditional assessment relied on ground surveys, windshield inspections, and reconnaissance teams, which delivered detailed building-level information but required substantial time, labor, and physical access to affected areas [44, 45]. While important for long-term model refinement and code calibration [46-48], these methods remain insufficient for real-time decision-making immediately after a hazard.

Remote sensing has expanded the achievable scale and timeliness of post-disaster assessments. Unmanned Aerial Vehicle (UAV) imagery offers fine spatial resolution for detailed structural evaluation, whereas aerial and satellite imagery provide rapid, large-area coverage essential for regional situational awareness [49]. In parallel, deep learning has become the dominant approach for automating damage inference from these imagery sources [50-53].

Beyond computer vision, other post-disaster data streams have emerged. Social sensing and outage reports offer indirect indicators of damage [54, 55], while crowdsourcing platforms provide rapid, human-generated observations at scale [56]. Although these sources broaden the available evidence, they also amplify challenges of heterogeneity, partial coverage, and variable fidelity.

Despite these advances, most post-disaster approaches remain fundamentally data-driven and lack grounding in physics-based vulnerability principles or statistically coherent inference. Some recent studies attempt to model uncertainty, such as uncertainty-aware crowdsourcing frameworks [57] and uncertainty-aware deep learning models for aerial-image classification [58], but these methods quantify uncertainty within their respective data sources rather than linking observations to prior structural vulnerability or enabling principled Bayesian updating. They do not reconcile inconsistent evidence, propagate information into unobserved areas, or integrate multi-fidelity observations within a unified probabilistic framework. In practice, data arrives gradually, with inconsistent fidelity and substantial spatial gaps. These limitations underscore the need for methodologies that incrementally update pre-event priors, assimilate heterogeneous post-disaster observations, and quantify uncertainty in a coherent statistical structure.

### *1.4. Background on Efforts to Integrate Predictive Risk Assessment with Adaptive Data Assimilation and the Resulting Research Gap*

Recent efforts have attempted to bridge pre-event predictive modeling with post-disaster data assimilation, but existing approaches remain fragmented and limited in scope. Some work has focused on adaptive data collection, such as the strategy introduced by Behrooz, Ilbeigi [59], which employs an ordinal gradient-boosting model and a multi-objective sampling scheme to prioritize informative buildings. However, these approaches rely on heuristic pre-event predictors and lack statistically coherent update mechanisms anchored in prior vulnerability functions. Other advances explore reinforcement learning to guide autonomous robotics for reconnaissance and search-and-rescue [60], though these efforts emphasize operational efficiency rather than principled updating of structural performance predictions.

Complementary research has incorporated post-disaster observations into broader risk and recovery frameworks. For example, Braik and Koliou [61] combined remote sensing, GIS, and deep learning to automate post-event damage assessment and used the resulting outputs for recovery forecasting. Building on this, Braik, Han [62] used computer-vision–derived damage states to initialize agent-based recovery models and to back-infer tornado wind fields. While valuable, these approaches assume timely, high-fidelity, and spatially complete observations—conditions rarely met immediately after a disaster—and do not provide a unified statistical foundation for dynamically updating fragility-based predictions.

Despite these advances, the integration of physics-informed fragility priors with incrementally arriving heterogeneous post-disaster evidence, reliability-dependent uncertainty, and spatial propagation remains underdeveloped. In particular, there remains a need for mathematically stable and computationally efficient structures that can sequentially assimilate probabilistic evidence while preserving valid multistate fragility estimates. This need is especially important in time-sensitive post-disaster settings, where inference must remain tractable, uncertainty-aware, and reliable despite incomplete and unevenly reliable observations. The present study addresses this gap through a two-stage Bayesian framework that combines analytically tractable local updating with ordered heteroscedastic spatial propagation. While hierarchical Bayesian formulations have been explored [63, 64], these efforts have largely focused on parameter estimation and have not addressed sequential updating, spatial propagation of uncertainty, or fusion of mixed-quality evidence.

A second major gap concerns the global propagation of local information. For interconnected systems such as electric power networks, Bayesian network–based digital twins have demonstrated how knowledge acquired at one component can inform others through principled dependency structures [65, 66]. However, structural building damage is inherently spatial rather than network-based, requiring models that encode proximity, correlation decay, and cross-archetype relationships. Recently, Braik and Koliou [67] proposed a spatial fragility modeling framework based on a warped Gaussian Process (GP) that represents fragility as a coherent field across space, archetypes, and damage states. While this approach establishes a strong mathematical foundation for spatial inference, its deployment for sequential updating and sequential data assimilation remains underexplored.

Collectively, these limitations reveal a critical research gap: there is no unified, mathematically coherent, dynamically updatable digital-twin framework that links pre-event fragility models with evolving post-disaster observations, propagates information spatially, and accommodates incomplete, heterogeneous, multi-fidelity data. Developing such a framework is essential for maintaining robust risk assessments before, during, and after a hazard event.

## 2. Methodology

The proposed methodology develops a sequential Bayesian learning framework for fragility modeling that continually refines physics-based predictions as new observations become available. It combines two complementary stages: a local updating stage that assimilates heterogeneous evidence to correct physics-based fragility estimates, and a global propagation stage that uses a structured GP to generalize these corrections across space and archetypes. This produces a coherent, end-to-end spatial learning pipeline that updates fragility estimates where direct evidence is available and transfers information to unobserved locations via kernel learning and hyperparameter tuning.

The framework begins by redefining conventional physics-based fragility functions as low-fidelity Probit–Normal (PN) priors. This anchors the learning process in established physics and keeps inference physically consistent even when data are sparse. The PN form captures both aleatory variability and epistemic uncertainty in an analytically tractable latent fragility index, providing a coherent baseline from which data-driven updates can evolve. In this way, the model first learns from physics and then from data, preserving interpretability and stability during early information scarcity.

The **first Bayesian stage** performs local updating of these PN priors. Each PN marginal is moment-matched to a Beta surrogate, enabling conjugate inference without the need for Markov chain Monte Carlo (MCMC) simulation. This mapping preserves the mean and variance of the PN distribution while maintaining statistical fidelity to its shape. Heterogeneous data sources are then assimilated through soft exceedance probabilities, with reliability weights encoding epistemic confidence. The resulting Beta posteriors are projected back into PN form, producing fragility observations whose variances reflect data fidelity—large for prior-only or uncertain sites and small where high-quality evidence is available. In this way, the model evolves from physics-based pseudo-observations toward increasingly data-driven fragility estimates as information accumulates.

The **second Bayesian stage** enables global learning and spatial generalization. It employs an ordered GP defined on a shared latent severity field. The model propagates information across space and archetypes while enforcing damage-state ordering through strictly positive threshold-gap functions. Thus, valid ordering is guaranteed not only for posterior mean exceedance probabilities but for every realization of the latent posterior field, conditional on the fitted gap-surface parameters.

Together, the two Bayesian stages form a sequential probabilistic learning system: local updates incorporate incoming evidence at observed buildings, while the ordered GP propagates information to unobserved or high-uncertainty buildings. This transforms fragility modeling from a static pre-event prediction into an event-conditioned spatial posterior field that evolves as evidence becomes available.

The overall workflow is summarized in Figure 1, with details provided in §2.1–2.3.

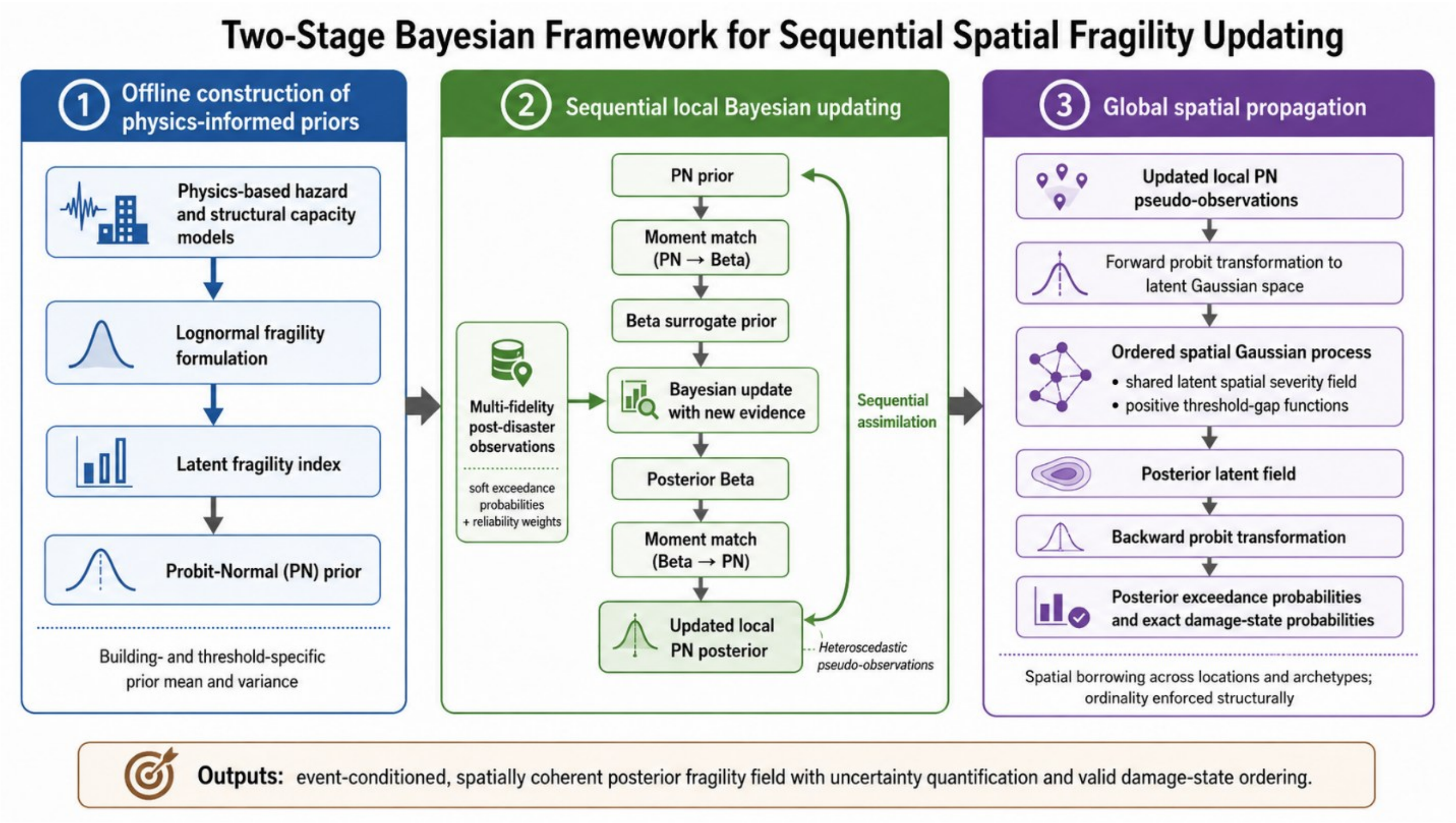

Figure 1: Schematic of the proposed two-stage Bayesian framework for sequential spatial fragility updating.

The proposed framework was evaluated through two complementary studies. First, a controlled synthetic validation study was conducted in §3 under known ground truth to assess the accuracy of local PN–Beta–PN updating, heteroscedastic spatial propagation, posterior uncertainty, and ordinal validity. Second, the framework was applied to the 2011 Joplin tornado in §4 to evaluate its scalability and practical behavior for a regional building inventory under progressively available post-event computer-vision evidence. The synthetic study provides a direct test of inferential validity because the true latent fragility field, threshold-specific exceedance probabilities, evidence reliability, and spatial information gaps are prescribed. The Joplin study then examines the operational behavior of the same framework under a realistic large-scale post-disaster setting.

### *2.1. Defining Physics-Informed Fragility Priors in Probit-Normal Form*

In the present framework, physics-based fragility functions are employed as low-fidelity priors for exceedance probabilities. Let $N$ denote the number of buildings in the inventory, $J$ the number of ordinal exceedance damage thresholds, and $A$ the number of structural archetypes. For building $i \in \{1, \ldots, N\}$, let $s_i \in \mathbb{R}^2$ denote its spatial location and let $a_i \in \{1, \ldots, A\}$ denote its structural archetype. For each ordered damage threshold $d_j$, with $j \in \{1, \ldots, J\}$, the probability of exceedance is expressed as:

$$P_{ij} \coloneqq \Pr\left(D_i \geq d_j | H_i, C_{ij}\right) = \Phi\left(\left(\ln(H_i) - \ln\left(C_{ij}\right)\right) / \beta_{ij}\right) \in (0,1) \qquad (1)$$

where $\Phi(\cdot)$ denotes the standard normal cumulative distribution function (CDF), $H_i$ is a random variable representing the hazard intensity at the location $s_i$, $C_{ij}$ is a random variable representing the epistemically uncertain median capacity threshold corresponding to the damage state $d_j$ for building $i$, and $\beta_{ij}$ is the logarithmic standard deviation (dispersion) of the fragility function for damage state $d_j$ of building $i$, reflecting aleatory variability in the demand-to-capacity ratio;

Equation (1) represents the standard lognormal fragility formulation commonly used in both seismic and wind engineering applications. To extend this formulation into a low-fidelity physics-informed prior, epistemic uncertainties in the hazard and capacity terms are included. These uncertainties are treated as lognormal random variables:

$$\begin{aligned} \ln(H_i) &\sim \mathcal{N}\left(\lambda_{H,i}\,,\, \beta_{H,i}^2\right) \\ \ln\left(C_{ij}\right) &\sim \mathcal{N}\left(\lambda_{C,ij}\,,\, \beta_{C,ij}^2\right) \end{aligned} \qquad (2)$$

where $\lambda_{H,i}\,, \beta_{H,i}$ denote the mean and standard deviation of $\ln(H_i)$, while $\lambda_{C,ij}\,, \beta_{C,ij}$ denote the mean and standard deviation of $\ln\left(C_{ij}\right)$. Hence, $\beta_{H,i}$ and $\beta_{C,ij}$ capture epistemic uncertainties associated with hazard estimation and damage state threshold, respectively.

A latent fragility index is defined by applying the probit transform to the exceedance probability:

$$Z_{ij} \coloneqq \Phi^{-1}\left(P_{ij}\right) = \left(\ln(H_i) - \ln\left(C_{ij}\right)\right) / \beta_{ij} \qquad (3)$$

Given the assumptions above, the latent variable follows a normal distribution:

$$Z_{ij} \sim \mathcal{N}\left(\mu_{ij}\,,\,\sigma_{ij}^2\right) \qquad (4)$$

with mean and variance

$$\begin{aligned} \mu_{ij} &= \left(\lambda_{H,i} - \lambda_{C,ij}\right)/\beta_{ij} \\ \sigma_{ij}^2 &= \left(\beta_{H,i}^2 + \beta_{C,ij}^2\right)/\beta_{ij}^2 \end{aligned} \qquad (5)$$

Thus, the exceedance probability becomes a bounded random variable through the probit transformation of the Gaussian latent index:

$$P_{ij} = \Phi\left(Z_{ij}\right) \sim \mathcal{PN}\left(\mu_{ij}\,,\,\sigma_{ij}^2\right) \in (0,1) \qquad (6)$$

The resulting exceedance probabilities $\{P_{ij}\}$ therefore follow a Probit–Normal (PN) distribution, for which closed-form expressions for the mean and variance are available [67]:

$$\begin{aligned} m_{ij} &\coloneqq \mathrm{E}\left[P_{ij}\right] = \Phi\left(\nu_{ij}\right) \\ \zeta_{ij} &\coloneqq \mathrm{Var}\left[P_{ij}\right] = \Phi_2\left(\nu_{ij}, \nu_{ij}, \eta_{ij}\right) - \Phi\left(\nu_{ij}\right)^2 \end{aligned} \qquad (7)$$

where:

$$\begin{aligned} \nu_{ij} &:= \mu_{ij}\,/\sqrt{1 + \sigma_{ij}^2} \\ \eta_{ij} &:= \sigma_{ij}^2/\left(1 + \sigma_{ij}^2\right) \end{aligned} \qquad (8)$$

### *2.2. First Bayesian Stage: Local Updating with Multi-Fidelity Observations*

#### *2.2.1. Moment Matching Between Probit–Normal and Beta Distributions*

While the PN distribution provides the canonical representation of exceedance probabilities derived from physics-based fragility (§2.1), it lacks the conjugacy required for efficient Bayesian updating with Bernoulli-type observations. By contrast, the Beta distribution is the natural conjugate prior for binomial and Bernoulli likelihoods [68], and therefore serves as an ideal vehicle for analytic data assimilation. To bridge these two representations, a moment-matched Beta surrogate is introduced.

A Beta distribution $\mathcal{B}(\alpha_{ij}, \gamma_{ij})$ is defined by equating its first two moments to $(m_{ij}, \zeta_{ij})$ of the PN distribution defined earlier in §2.1:

$$\begin{aligned} \alpha_{ij} &= m_{ij}\varpi_{ij} \\ \gamma_{ij} &= (1 - m_{ij})\varpi_{ij} \end{aligned} \qquad (9)$$

where:

$$\varpi_{ij} = m_{ij}(1 - m_{ij})/\zeta_{ij} - 1 \qquad (10)$$

The feasibility is guaranteed by the inequality $0 < \zeta_{ij} < m_{ij}(1 - m_{ij})$, which is always satisfied for PN marginals.

This construction is more than heuristic. The Beta surrogate preserves the exact mean and variance of the PN distribution while simultaneously enabling analytic conjugacy. Beyond the moment match, the two distributions are very similar in shape. Figure 2 compares PN and Beta densities across representative $\mu$ and $\sigma^2$ values. The match is nearly indistinguishable across the operational domain of interest, with discrepancies appearing only in the extreme tails, where probabilities approach 0 or 1.

The information loss incurred by substituting PN with its moment-matched Beta surrogate is quantified using the Kullback–Leibler divergence ($D_{KL}$) [69]. Figure 3 shows heatmaps of $D_{KL}(\text{PN}||\text{Beta})$ and $D_{KL}(\text{Beta}||\text{PN})$ over $\mu \in [-3,3]$ and $\sigma^2 \leq 3$. Two insights emerge. First, across the entire range, $D_{KL}$ remains below 0.11, which corresponds to less than 0.11 nats of expected excess information loss per update, which is considered negligible in the information-theoretic interpretation of $D_{KL}$ [69, 70]. Second, the regions of largest $D_{KL}$ correspond to cases where $|\mu|$ is large, and $\sigma^2$ is small, i.e., where the fragility distribution is nearly deterministic with probabilities pinned near 0 or 1. In these regimes, the posterior updates are effectively deterministic because both PN and Beta yield near-zero variance, and hence, the exact shape of the PDF is of little engineering consequence. In the transition regions, where uncertainty is greatest and engineering decisions are most sensitive, $D_{KL}$ is almost zero, indicating that the Beta surrogate is an almost exact proxy for PN.

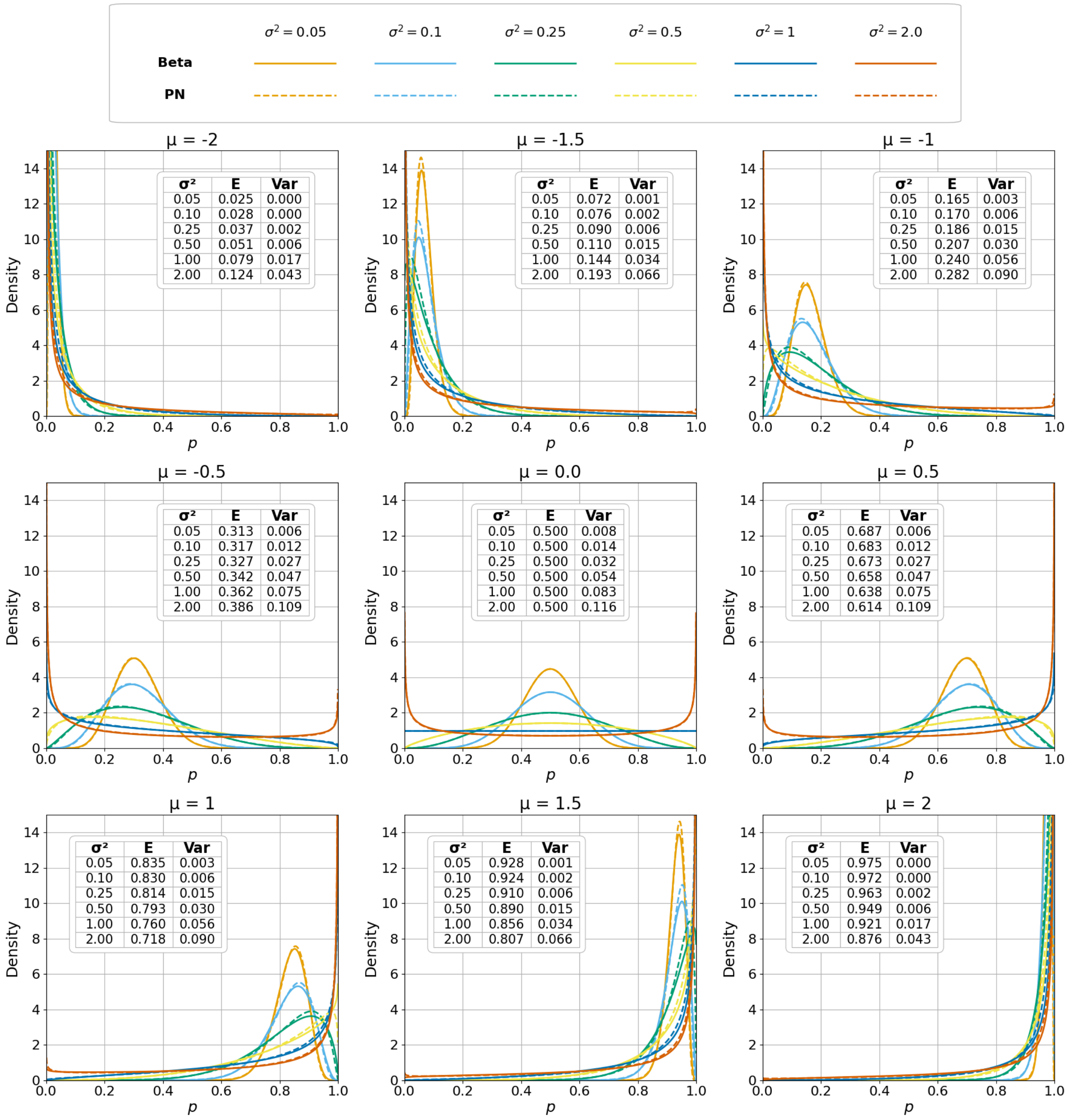

Figure 2: Comparison of PN and Beta Distributions Across Representative Parameters

Accordingly, the moment-matched Beta distribution provides a principled and empirically validated surrogate for PN marginals: it retains exact moment information, incurs negligible information loss, and enables conjugate Bayesian updating. Within the framework, the PN distribution remains the canonical representation for spatial GP modeling (§2.3), while the PN–Beta–PN pipeline serves as the computational mechanism for efficient sequential data assimilation (§2.2.2). The numerical results in Appendix A confirm that, after evidence assimilation and reprojection to the PN representation, the resulting local posterior approximation remains close to the converged quadrature posterior over the principal operational domain while substantially reducing repeated-update computation.

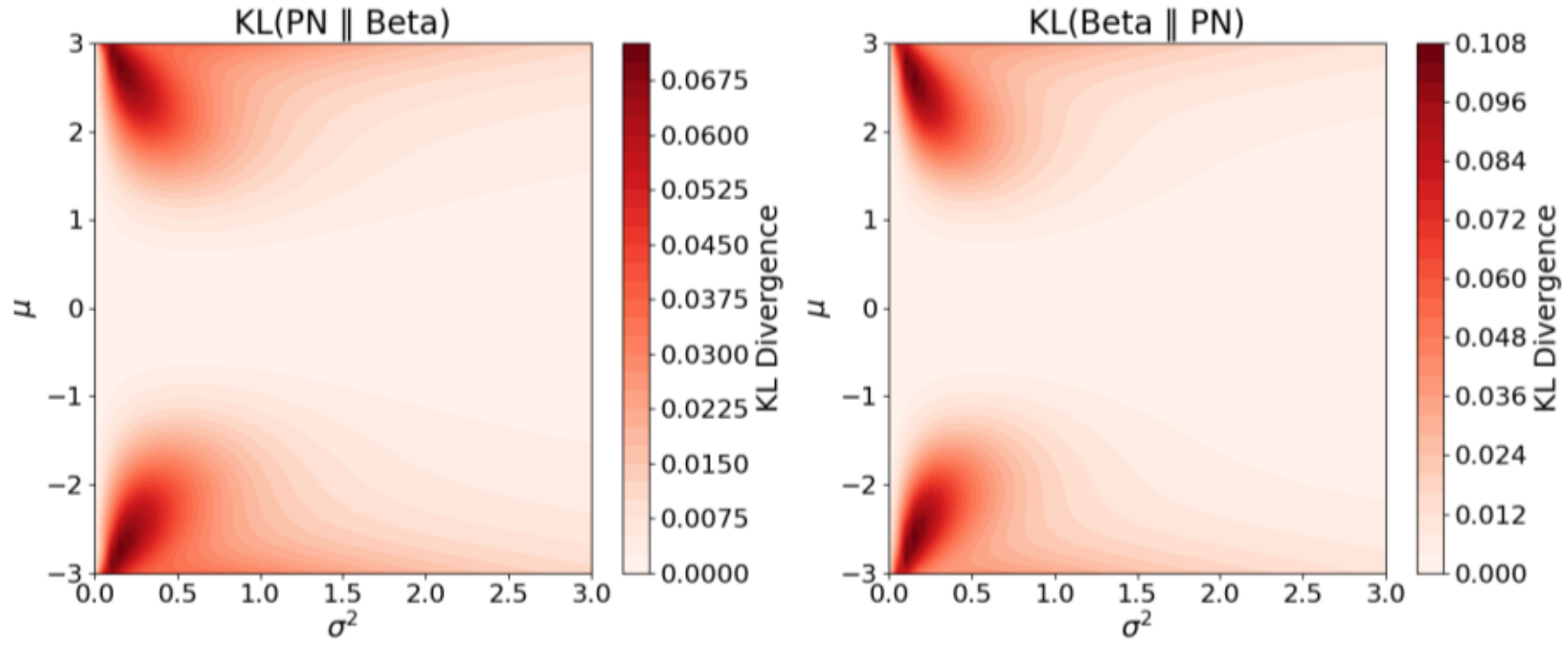

Figure 3: KL Divergence Between PN and Beta Distributions Over Parameter Space

### *2.2.2. Bayesian Updating with Multi-Fidelity Observations*

The moment-matched Beta surrogate introduced in §2.2.1 enables closed-form Bayesian updating when new evidence becomes available. In practice, observations may be derived from diverse sources such as satellite and aerial imagery processed by machine learning classifiers [53], rapid visual inspections [71], or sensor-based measurements [72]. These sources vary in reliability and resolution and introduce two distinct forms of uncertainty. The first is aleatory variability, which is captured by the probabilistic outputs of the data source. The second is epistemic uncertainty, which reflects the trustworthiness of the source and is encoded as a weight factor calibrated using independent validation data.

For building $i$ and damage exceedance threshold $d_j$, source $b$ provides a probabilistic statement of exceedance denoted by $y_{ij}^{(b)} \in (0,1)$. This represents the aleatory uncertainty in whether $D_i \geq$ $d_j$. In cases where the source outputs categorical probabilities $\left\{\Im_{i\chi}^{(b)}\right\}_{\chi=0}^{J}$ for the $J+1$ ordered damage classes (including the no damage class), the exceedance probability is obtained by cumulative aggregation:

$$y_{ij}^{(b)} = \sum_{\chi=j}^{J} \Im_{i\chi}^{(b)} \qquad (11)$$

Thus, $y_{ij}^{(b)}$ constitutes a soft observation of the exceedance event, directly encoding the variability conveyed by the source, while the epistemic reliability of each source is captured by a nonnegative weight $w_j^{(b)}$. In this manner, high-performing sources exert greater influence on the posterior, while less reliable sources are naturally down-weighted.

Given the Beta prior $\mathcal{B}(\alpha_{ij}, \gamma_{ij})$, the Bayesian update rule is expressed as [73]:

$$\begin{aligned} \alpha'_{ij} &= \alpha_{ij} + \sum_{b=1}^{B} w_j^{(b)} y_{ij}^{(b)} \\ \gamma'_{ij} &= \gamma_{ij} + \sum_{b=1}^{B} w_j^{(b)} \left(1 - y_{ij}^{(b)}\right) \end{aligned} \qquad (12)$$

with posterior distribution:

$$P_{ij}|\text{data} \sim \mathcal{B}(\alpha'_{ij}, \gamma'_{ij}) \qquad (13)$$

Posterior moments follow directly:

$$\begin{aligned} m'_{ij} &= \alpha'_{ij} / (\alpha'_{ij} + \gamma'_{ij}) \\ \zeta'_{ij} &= \alpha'_{ij}\gamma'_{ij} / \left((\alpha'_{ij} + \gamma'_{ij})^2 (\alpha'_{ij} + \gamma'_{ij} + 1)\right) \end{aligned} \qquad (14)$$

Because the Beta update is associative and commutative, evidence from multiple sources or batches may be assimilated in any order. In this way, physics-based fragility priors are systematically refined by heterogeneous data streams, producing posterior distributions that coherently integrate structural knowledge and observational fidelity.

Figure 4 illustrates the first Bayesian updating stage. The PN prior derived from physics-based fragility (§2.1) is first mapped to a moment-matched Beta prior (§2.2.1) to enable conjugate updating with soft evidence from heterogeneous observation sources. After incorporating multi-fidelity observations (§2.2.2), the resulting Beta posterior is mapped back to the PN representation. This restores a latent fragility index form that remains fully compatible with the broader spatial and multi-output modeling framework introduced in §2.3. This cyclic mapping defines the local Bayesian update stage.

$$\text{PN prior} \xrightarrow[\text{moment match}]{} \text{Beta prior} \xrightarrow[\text{soft evidence}]{} \text{Beta posterior} \xrightarrow[\text{moment match back}]{} \text{PN posterior}$$

Figure 4: Local Bayesian Updating Cycle (PN–Beta–PN)

### 2.2.2.1. *Calibration of Epistemic Weights*

The epistemic weight $w_j^{(b)}$ introduced above quantifies the reliability of source $b$ for predicting exceedance at threshold $d_j$. Whereas the soft exceedance probabilities $y_{ij}^{(b)}$ encode aleatory variability for each building $i$ in the application dataset, the weights $w_j^{(b)}$ are derived from a designated evaluation subset of the application data. This ensures that the reliability factors reflect the observed performance of the source in the same operational domain in which the probabilistic scores are applied.

As an example, when source $b$ is a machine learning model employed for computer-vision–based damage assessment, its reliability is quantified on labeled evaluation imagery. For each exceedance threshold $d_j$, $o_{\varsigma j} \in \{0,1\}$ denote the ground-truth indicator that sample $\varsigma$ from the evaluation subset satisfies $D \geq d_j$, while $g_{\varsigma j}^{(b)} \in (0,1)$ denote the predicted exceedance probability produced by source $b$ for that event. Hence, the soft exceedance counts across evaluation samples $\varsigma$ are:

$$
\begin{aligned}
TP_j^{(b)} &= \sum_{\varsigma} g_{\varsigma j}^{(b)}\, o_{\varsigma j} \\
FP_j^{(b)} &= \sum_{\varsigma} g_{\varsigma j}^{(b)} \left(1 - o_{\varsigma j}\right) \\
FN_j^{(b)} &= \sum_{\varsigma} \left(1 - g_{\varsigma j}^{(b)}\right) o_{\varsigma j}
\end{aligned}
\qquad (15)
$$

where $TP$, $FP$, and $FN$ denote the soft analogues of true positives, false positives, and false negatives, respectively. The resulting soft exceedance F1 score is defined as:

$$
F1_j^{(b)} = 2TP_j^{(b)} / \left(2TP_j^{(b)} + FP_j^{(b)} + FN_j^{(b)}\right) \qquad (16)
$$

Finally, the soft F1 score is transformed into an epistemic reliability weight that ensures non-negativity, grows monotonically with predictive skill, and increases logarithmically as $F1_j^{(b)} \rightarrow 1$:

$$
w_j^{(b)} = \log_2 \left(1 / \left(1 - F1_j^{(b)}\right)^2\right) \qquad (17)
$$

The transformation uses a squared gap $\left(1 - F1_j^{(b)}\right)^2$ to impose a symmetric penalty on both precision and recall, consistent with the definition of the F1 score, and to produce a stricter separation between moderately accurate and highly accurate sources. Squaring also reduces the sensitivity of the reliability weight to sampling noise in F1 estimates. A base-2 logarithm is adopted, yielding the interpretable relationship $1 - F1 = 2^{-w/2}$, whereby each +2 units of weight correspond to halving the remaining prediction error.

### *2.3. Second Bayesian Stage: Global Spatial Propagation*

Once local fragility indices are updated through the PN–Beta moment-matching procedure described in §2.2, the resulting estimates reflect building- and threshold-specific evidence with variances that may differ substantially. This heterogeneity arises because some buildings may receive repeated or high-quality observations, whereas others may remain prior dominated or be observed only through lower-reliability evidence. Consequently, some pseudo-observations have relatively small uncertainty, while others remain highly uncertain.

The combination of this heteroscedasticity with the Gaussian structure induced by the probit transformation makes GPs a natural framework for the second stage of inference. The GP accounts for differing pseudo-observation reliabilities through their associated variances and propagates information across space and archetypes. To ensure valid ordinal exceedance probabilities, the present formulation uses a shared latent spatial severity field together with strictly positive threshold-gap functions.

### *2.3.1. Spatial Fragility Modeling Using an Ordered Probit-Warped Gaussian Process*

Let $x_i$ denote the vector of building-level covariates for building $i$. Let $F_i = F(s_i, a_i)$ denote the shared latent spatial severity field for building $i$. For $J$ ordered exceedance thresholds, the threshold-specific latent fragility indices are defined as:

$$Z_{ij} = F_i - \sum_{t=1}^{j-1} \Delta_{t,t+1}(x_i), \qquad j = 2, \dots, J \tag{18}$$

where $Z_{i1} = F_i$, and $\Delta_{\mathrm{t,t+1}}(x_i)$ denotes the positive latent separation between adjacent exceedance thresholds $t$ and $t + 1$. Each adjacent threshold gap is parameterized as:

$$\Delta_{t,t+1}(x_i) = \mathrm{softplus}\left[h_{t,t+1}(x_i)^{\top}\theta_{t,t+1}\right], \qquad t = 1, \dots, J - 1 \tag{19}$$

where:

$$\mathrm{softplus}(u) = \log(1 + exp(u)) \tag{20}$$

Because the softplus transformation is strictly positive, $\Delta_{t,t+1}(x_i) > 0$ for all $t = 1, \dots, J - 1$, therefore, $Z_{i1} > Z_{i2} > \cdots > Z_{iJ}$ for every building and every realization of the shared latent field $F_i$, conditional on the fitted gap-surface parameters. Since $\Phi(\cdot)$ is strictly increasing, $P_{i1} > P_{i2} > \cdots > P_{iJ}$ also holds pathwise.

The shared latent severity field is assigned a GP prior:

$$F(s,a) \sim \mathcal{GP}\left[m_F(s,a), k_F\big((s,a),(s',a')\big)\right] \qquad (21)$$

The covariance function combines spatial and archetype dependence:

$$k_F[(s,a),(s',a')] = \sigma_F^2 k_{\text{sp}}(s,s';\ell)k_{\text{arch}}(a,a';\rho_a) \qquad (22)$$

where $\sigma_F^2 > 0$ is the latent-field variance, $\ell$ denotes spatial lengthscale parameters, and $\rho_a$ controls cross-archetype dependence.

The anisotropic squared-exponential spatial kernel is

$$k_{\text{sp}}(s,s') = \exp\left[-\frac{1}{2}\left(\frac{s_1 - s_1'}{\ell_1}\right)^2 - \frac{1}{2}\left(\frac{s_2 - s_2'}{\ell_2}\right)^2\right] \qquad (23)$$

The exchangeable archetype kernel is

$$k_{\text{arch}}(a,a';\rho_a) = \begin{cases} 1, & a = a', \\ \rho_a, & a \neq a', \end{cases} \quad 0 \leq \rho_a < 1 \qquad (24)$$

This kernel permits spatial dependence to decay differently along the two coordinate directions while allowing partial information transfer among distinct structural archetypes. The restriction $0 \leq \rho_a < 1$ is imposed intentionally to permit nonnegative information sharing across distinct archetypes while preserving a positive-semidefinite exchangeable covariance structure.

The shared-field construction intentionally restricts threshold-specific stochastic flexibility. All exceedance thresholds share one stochastic spatial severity field, whereas their separation is governed by deterministic positive gap surfaces. This restriction trades unrestricted threshold-specific spatial deviations for pathwise ordinal validity, nonnegative exact-state probabilities, and a coherent multistate damage representation.

The physics-informed fragility prior represents broad inventory archetypes and therefore cannot capture all building-specific characteristics that influence realized damage. The spatial GP is introduced to learn residual dependence from evidence-updated pseudo-observations rather than assuming that nearby buildings or those within the same archetype perform identically. The degree of information sharing is estimated from the fitted covariance structure and moderated by the local posterior and discrepancy variances, thereby allowing uncertainty to remain elevated where building-specific variability is not explained by the available predictors.

### *2.3.2. Ordered Heteroscedastic Pseudo-Observation Model*

The Stage-1 posterior means on the probit scale are used as modular pseudo-observations of the ordered latent fragility indices. Conditional on the Stage-1 analysis, these pseudo-observations are modeled as:

$$\mu_{ij,\mathrm{loc}} = F_i - \sum_{t=1}^{j-1} \Delta_{t,t+1}(x_i) + \varepsilon_{ij}, \qquad j = 2, \dots, J \tag{25}$$

where $\mu_{i1,\mathrm{loc}} = F_i + \varepsilon_{i1}$. The residual terms satisfy:

$$\varepsilon_{ij} \sim \mathcal{N}\left(0, \sigma_{ij,\mathrm{loc}}^2 + \tau_j^2\right) \tag{26}$$

where $\sigma_{ij,\mathrm{loc}}^2$ is the local PN variance obtained from Stage 1 and $\tau_j^2 \geq 0$ is a threshold-specific discrepancy variance. The discrepancy term accounts for residual mismatch between the PN–Beta–PN approximation and the ordered spatial model, including possible underestimation of uncertainty by the Stage-1 pseudo-observation variance. The observation model is therefore heteroscedastic at both building and threshold levels. Buildings with more reliable local evidence receive greater likelihood precision, while high-uncertainty buildings borrow more information from spatially close or structurally similar buildings.

Stacking all threshold-specific pseudo-observations across buildings gives:

$$z_{\mathrm{loc}} = \mathrm{vector}\left\{\mu_{ij,loc}\right\}_{(i,j)\in O} = \mathbf{A}F + g(\theta) + \varepsilon \tag{27}$$

where $F$ is the vector of shared latent-field values, $\mathbf{A}$ maps the shared field to the stacked threshold observations, $g(\theta)$ contains the negative threshold-gap contributions, and $\varepsilon$ is the stacked residual vector. Conditional on the gap-surface parameters and GP hyperparameters:

$$z_{\text{loc}} \sim \mathcal{N}(\mathbf{A}m_F + g(\theta), \mathbf{K}_{\text{obs}}) \qquad (28)$$

where:

$$\mathbf{K}_{\text{obs}} = \mathbf{A}\mathbf{K}_F\mathbf{A}^\top + \mathbf{\Sigma}_{\text{loc}} \qquad (29)$$

and:

$$\mathbf{\Sigma}_{\text{loc}} = \text{diag}\left[\sigma_{ij,\text{loc}}^2 + \tau_j^2\right] \qquad (30)$$

The likelihood retains all threshold-specific pseudo-observations. It does not collapse the $J$ threshold observations for a building into one aggregate quantity because their within-building contrasts provide information needed to estimate the threshold-gap surfaces. For moderate systems, parameters may be estimated by maximizing the exact marginal likelihood [74]:

$$\log p(z_{loc} \mid \theta, \psi) = -\frac{1}{2}r^\top\mathbf{K}_{\text{obs}}^{-1}r - \frac{1}{2}\log|\mathbf{K}_{\text{obs}}| - \frac{n_{\text{obs}}}{2}\log(2\pi) \qquad (31)$$

where $\psi = \left(\sigma_F^2, \ell_{\text{lat}}, \ell_{\text{lon}}, \rho_a, \tau_1^2, \dots, \tau_J^2\right)$, and:

$$r = z_{\text{loc}} - \mathbf{A}m_F - g(\theta) \qquad (32)$$

Under the exact GP formulation, the posterior distribution of the shared latent field at prediction locations is [75]:

$$F_* \mid z_{\text{loc}} \sim \mathcal{N}\left(\mu_{F,*}, \mathbf{\Sigma}_{F,*}\right) \qquad (33)$$

with:

$$\mu_{F,*} = m_{F,*} + \mathbf{K}_{*,\text{obs}}\mathbf{K}_{\text{obs}}^{-1}r \qquad (34)$$

and:

$$\mathbf{\Sigma}_{F,*} = \mathbf{K}_{**} - \mathbf{K}_{*,\text{obs}}\mathbf{K}_{\text{obs}}^{-1}\mathbf{K}_{\text{obs},*} \qquad (35)$$

where $\mathbf{K}_{*,\text{obs}} = \text{Cov}(F_*, z_{\text{loc}}) = \mathbf{K}_{F,*}\mathbf{A}^\top$ , $\mathbf{K}_{\text{obs},*} = \mathbf{K}_{*,\text{obs}}^\top = \mathbf{A}\mathbf{K}_{F,*}^\top$ and $\mathbf{K}_{F,*}$ denotes the covariance between the prediction locations and the building-level latent-field values.

For regional inventories containing many thousands of buildings, exact inference becomes computationally impractical because direct matrix operations scale cubically with the number of stacked pseudo-observations. Therefore, sparse variational GP inference is adopted [76, 77].

Let $u$ denote inducing variables at inducing inputs $X_u$, with variational distribution:

$$q(u) = \mathcal{N}(m_u, \mathbf{S}_u) \qquad (36)$$

The evidence lower bound (ELBO) is

$$\mathcal{L} = \sum_{(i,j)\in O} \mathbb{E}_{q(F_i)} \left[\log\left(p\left(\mu_{ij,\text{loc}} \mid F_i, \theta, \sigma_{ij,\text{loc}}^2, \tau_j^2\right)\right)\right] - \text{KL}[q(u) \parallel p(u)] \qquad (37)$$

A weak zero-centered Gaussian prior is imposed on the threshold-gap coefficients, which is equivalent to maximizing the penalized ELBO:

$$\mathcal{L}_{\text{MAP}} = \mathcal{L} - \lambda_{\text{gap}} \sum_{t=1}^{J-1} \parallel \theta_{t,t+1} \parallel_2^2 \qquad (38)$$

This weak regularization stabilizes estimation of the gap surfaces but does not impose ordinality, which is guaranteed by the softplus parameterization. Conditional on the inducing inputs, the covariance parameters, archetype-correlation parameter, variational parameters, threshold-gap coefficients, and discrepancy variances are estimated jointly by maximizing the exact marginal likelihood or penalized ELBO, as applicable.

### *2.3.3. Posterior Exceedance and Exact-State Probabilities*

Let $\mathcal{D}$ denote the Stage-2 data set, consisting of the locally updated probit-scale pseudo-observation means and variances, together with the associated building locations, structural archetypes, and covariates, for all observed building–threshold pairs. After fitting the ordered GP, the posterior distribution of the shared latent field at building $i$ is:

$$F_i \mid \mathcal{D} \sim \mathcal{N}\left(\mu_{F,i}^{\star}, \sigma_{F,i}^{2\star}\right) \qquad (39)$$

where * denotes prediction-location posterior quantities. Conditional on the fitted gap-surface parameters:

$$Z_{i1} \mid \mathcal{D} \sim \mathcal{N}\left(\mu_{F,i}^{\star}, \sigma_{F,i}^{2\star}\right) \qquad (40)$$

and, for $j = 2, \dots, J$:

$$Z_{ij} \mid \mathcal{D} \sim \mathcal{N}\left[\mu_{F,i}^{\star} - \sum_{t=1}^{j-1} \Delta_{t,t+1}(x_i), \sigma_{F,i}^{2\star}\right] \qquad (41)$$

The threshold-specific latent indices are not conditionally independent. Because all thresholds are generated from the same posterior field $F_i$, their conditional covariance satisfies $\mathrm{Cov}\left(Z_{ij}, Z_{ik} \mid \mathcal{D}\right) = \sigma_{F,i}^{2\star}$ for all $j, k$. Conditional on the fitted gap surfaces, the latent threshold-index vector $\left(Z_{i1}, \dots, Z_{iJ}\right)^{\top}$ has covariance matrix $\sigma_{F,i}^{2\star}\mathbf{1}\mathbf{1}^{\top}$, which has rank one whenever $\sigma_{F,i}^{2\star} > 0$ and rank zero when $\sigma_{F,i}^{2\star} = 0$. This shared-field dependence preserves the pathwise ordering established earlier.

Define:

$$\mu_{ij}^{\star} = \mu_{F,i}^{\star} - \sum_{t=1}^{j-1} \Delta_{t,t+1}(x_i), \qquad (42)$$

where the summation is zero for $j = 1$. The posterior expected exceedance probability is [67]:

$$\mathbb{E}\left[P_{ij} \mid \mathcal{D}\right] = \Phi\left(\mu_{ij}^{\star} / \sqrt{1 + \sigma_{F,i}^{2\star}}\right) \qquad (43)$$

For compactness, let $c_{ij} = \sum_{t=1}^{j-1} \Delta_{t,t+1}(x_i)$, where $c_{i1} = 0$, so that $Z_{ij} = F_i - c_{ij}$. Conditional on the fitted gap surfaces, the exceedance probabilities are:

$$P_{ij} = \Phi(Z_{ij}) = \Phi(F_i - c_{ij}). \qquad (44)$$

Thus, each marginal exceedance probability follows a univariate Probit–Normal distribution. Let $q_{ij} = \mu_{ij}^{\star} / \sqrt{1 + \sigma_{F,i}^{2\star}}$. Its posterior mean is given by Eq. (43), and its marginal density is:

$$f_{P_{ij}|\mathcal{D}}(p) = \frac{\phi\left[\Phi^{-1}(p); \mu_{ij}^{\star}, \sigma_{F,i}^{2\star}\right]}{\phi[\Phi^{-1}(p)]}, 0 < p < 1, \qquad (45)$$

where $\phi(\cdot; \mu, \sigma^2)$ denotes the normal density with mean $\mu$ and variance $\sigma^2$, and $\phi(\cdot)$ denotes the standard normal density. For any two exceedance thresholds $j$ and $k$ of building $i$, define:

$$\rho_i^{\dagger} = \frac{\sigma_{F,i}^{2\star}}{1 + \sigma_{F,i}^{2\star}} \qquad (46)$$

The joint second moment is:

$$\mathbb{E}[P_{ij} P_{ik} \mid \mathcal{D}] = \Phi_2\left(q_{ij}, q_{ik}; \rho_i^{\dagger}\right), \qquad (47)$$

where $\Phi_2(\cdot,\cdot; \rho)$ is the standard bivariate normal CDF with correlation $\rho$. Therefore:

$$\mathrm{Cov}\left(P_{ij}, P_{ik} \mid \mathcal{D}\right) = \Phi_2\left(q_{ij}, q_{ik}; \rho_i^{\dagger}\right) - \Phi(q_{ij})\Phi(q_{ik}) \qquad (48)$$

In particular:

$$\mathrm{Var}(P_{ij} \mid \mathcal{D}) = \Phi_2\left(q_{ij}, q_{ij}; \rho_i^{\dagger}\right) - \Phi(q_{ij})^2 \qquad (49)$$

Although $\left(P_{i1}, \dots, P_{iJ}\right)$ is a degenerate multivariate Probit–Normal random vector, it does not possess a regular $J$-dimensional density with respect to Lebesgue measure. Conditional on the fitted gap surfaces, all components are generated by the single random variable $F_i$; therefore, the joint distribution is supported on the one-dimensional ordered curve $\left\{\left[\Phi(f - c_{i1}), \dots, \Phi(f - c_{iJ})\right] : f \in \mathbb{R}\right\}$. Its bivariate joint CDF for any pair of thresholds is nevertheless available in closed form:

$$\Pr\left(P_{ij} \le p_j,\ P_{ik} \le p_k \mid \mathcal{D}\right) = \Phi\left[\frac{\min\left\{\Phi^{-1}(p_j) + c_{ij}, \Phi^{-1}(p_k) + c_{ik}\right\} - \mu_{F,i}^{\star}}{\sigma_{F,i}^{\star}}\right] \qquad (50)$$

for $0 < p_j, p_k < 1$ and $\sigma_{F,i}^{\star} > 0$.

For general $J$, define exact damage-state probabilities by:

$$\pi_{ij} = P_{ij} - P_{i,j+1}, \qquad j = 1, \dots, J - 1 \tag{51}$$

where $\pi_{i0} = 1 - P_{i1}$ denotes the no exceedance state and $\pi_{iJ} = P_{iJ}$ denotes the most severe state. Because $1 \geq P_{i1} \geq P_{i2} \geq \cdots \geq P_{iJ} \geq 0$, all exact-state probabilities are nonnegative and sum to one for every realization of the latent field. Their posterior expectations follow directly by linearity:

$$\mathbb{E}\left[\pi_{ij} \mid \mathcal{D}\right] = \mathbb{E}\left[P_{ij} \mid \mathcal{D}\right] - \mathbb{E}\left[P_{i,j+1} \mid \mathcal{D}\right], \qquad j = 1, \dots, J - 1, \tag{52}$$

where $\mathbb{E}[\pi_{i0} \mid \mathcal{D}] = 1 - \mathbb{E}[P_{i1} \mid \mathcal{D}$ and $\mathbb{E}[\pi_{iJ} \mid \mathcal{D}] = \mathbb{E}[P_{iJ} \mid \mathcal{D}]$.

This differs fundamentally from fitting independent GPs to each threshold. Independent threshold-wise GPs can produce crossing posterior realizations even when posterior means appear ordered. Consequently, differencing independently fitted exceedance probabilities can generate negative exact damage-state probabilities. The present construction precludes this failure by enforcing ordering structurally through the shared latent severity field and strictly positive threshold-gap functions.

The ordering guarantee is conditional on fitted gap-surface parameters. In the present implementation, kernel and gap-surface parameters are estimated through exact marginal-likelihood optimization or sparse variational optimization. Full propagation of hyperparameter uncertainty would require a fully Bayesian treatment of these parameters and is a potential extension.

## 3. Controlled Synthetic Validation

A controlled synthetic study was conducted to evaluate the proposed two-stage framework under known latent fragility fields and known exceedance probabilities. The experiments were designed to assess five complementary properties: (i) covariance-supported recovery of an unobserved region as exterior evidence becomes denser; (ii) local PN–Beta–PN updating under evidence sources of differing calibrated reliability; (iii) recovery and uncertainty calibration of the fully fitted ordered GP in repeated contiguous-block holdouts; (iv) robustness of the complete end-to-end framework under a biased prior and noisy heterogeneous evidence; and (v) pathwise ordinal validity relative to unconstrained threshold-wise GP fits.

The synthetic inventory contained 169 buildings arranged on a regular spatial grid and assigned to three intermingled structural archetypes. Three ordered exceedance thresholds were considered. The true threshold-specific latent indices were generated from the ordered shared-field formulation of §2.3, consisting of a common latent spatial field and two positive adjacent threshold-gap surfaces. Thus, the true exceedance probabilities were ordered at every building by construction.

The first experiment isolated covariance-supported spatial propagation under the correctly specified ordered-GP observation model. Twenty-four independent latent-field realizations were generated. For each realization, a contiguous information-gap region received no pseudo-observations initially, while evidence outside the gap was introduced through six nested configurations: no evidence, sparse exterior evidence, moderate exterior evidence, dense boundary-ring evidence, all exterior evidence, and all exterior evidence supplemented with limited interior anchors.

Figure 5 summarizes recovery within the information-gap region. Stages 0–4 contain no pseudo-observations inside the gap. Nevertheless, increasing the amount and spatial connectivity of exterior evidence reduced the mean gap-region Brier score monotonically from 0.0591 with no evidence to 0.0380 under sparse exterior evidence, 0.0262 under moderate exterior evidence, 0.0213 under dense boundary-ring evidence, and 0.0127 after all exterior evidence was assimilated. The corresponding latent-field RMSE decreased from 0.9207 to 0.4320, while the mean posterior variance contracted from 0.9000 to 0.1829.

At the final stage, a limited number of direct anchors were introduced within the information gap. The mean Brier score, latent-field RMSE, and posterior variance then further decreased to 0.0018, 0.1487, and 0.0220, respectively. These results demonstrate that the ordered GP can recover the broad latent severity pattern in an unobserved contiguous region from increasingly informative surrounding evidence, while retaining greater uncertainty within the region until direct observations become available.

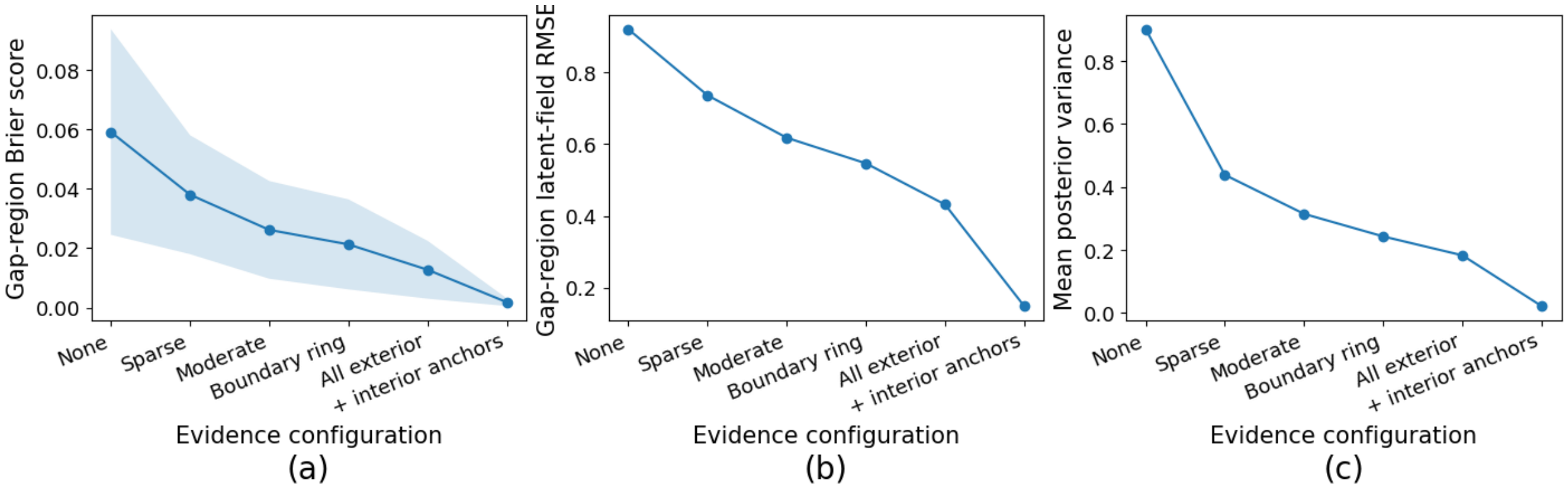

Figure 5: Model-consistent recovery of an unobserved spatial region under progressively denser exterior evidence. (a) Mean gap-region Brier score, with shaded bands denoting between-realization variability. (b) Mean latent-field RMSE within the information-gap region. (c) Mean posterior variance within the information-gap region.

The second experiment isolated the first Bayesian stage. A biased low-fidelity PN prior was generated with a smooth domain-wide discrepancy relative to the known exceedance probabilities. Low-, medium-, and high-fidelity evidence streams were then assimilated sequentially through the PN–Beta–PN update. Each source produced ordinal categorical evidence, which was transformed into threshold-specific exceedance information. The associated power-likelihood weights were calibrated from an independent evaluation subset using the threshold-specific soft F1 scores defined in §2.2.2.1.

The calibrated F1 scores increased systematically from low to medium to high fidelity across all thresholds. The corresponding weights also increased with source quality. For the Moderate,

Extensive, and Complete thresholds, respectively, the low-fidelity weights were 4.713, 4.224, and 3.257; the medium-fidelity weights were 6.994, 6.376, and 5.424; and the high-fidelity weights were 10.151, 9.410, and 8.552.

Figure 6(a) presents representative local posterior trajectories selected from prior-underprediction, transition-region, and prior-overprediction cases. The trajectories illustrate that individual posterior responses depend on the local relationship between prior error, realized evidence, and source reliability. Therefore, intermediate updates need not move monotonically toward the truth when evidence is noisy or partially conflicting.

Aggregate performance is summarized in Figure 6 and Table 1. Low-fidelity evidence alone slightly increased the Brier score from 0.0224 to 0.0231 and the threshold RMSE from 0.1497 to 0.1519. However, after adding medium-fidelity evidence, the Brier score decreased to 0.0185 and the threshold RMSE decreased to 0.1362. Assimilation of high-fidelity evidence further reduced these quantities to 0.0179 and 0.1337, respectively. Thus, reliability tempering limits, but does not guarantee elimination of, the influence of weak evidence; sufficiently informative medium- and high-fidelity evidence substantially improves the local posterior estimates.

Table 1. Stage-1 PN–Beta–PN performance under sequential evidence assimilation.

| **Evidence set** | **Brier** | **Log score** | **Threshold RMSE** |
|---|---|---|---|
| Prior only | 0.0224 | 0.4626 | 0.1497 |
| Low fidelity | 0.0231 | 0.4642 | 0.1519 |
| Low + medium fidelity | 0.0185 | 0.4534 | 0.1362 |
| Low + medium + high fidelity | 0.0179 | 0.4540 | 0.1337 |

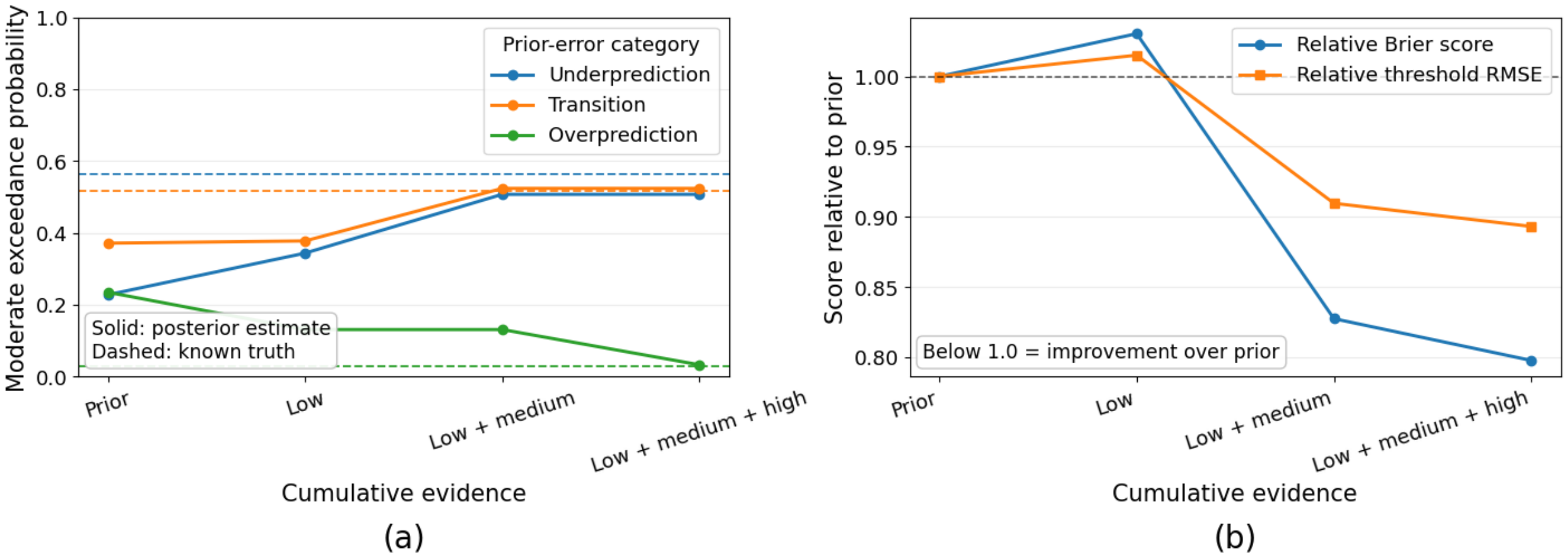

Figure 6: Local PN–Beta–PN updating with sequentially assimilated evidence of differing reliability. (a) Representative posterior trajectories for prior-underprediction, transition-region, and prior-overprediction cases. Solid lines denote posterior exceedance probabilities and matching dashed lines denote the known exceedance probabilities. (b) Aggregate Brier score and threshold RMSE relative to their prior values; values below one indicate improvement over the prior.

The third experiment evaluated recovery by the fully fitted ordered model in repeated contiguous-block holdouts. Four $5 \times 5$ blocks located in the southwest, southeast, northwest, and northeast portions of the domain were specified before generating the latent truth or pseudo-observations. Six independently generated realizations were evaluated for each block, producing 24 predeclared holdout evaluations.

Each held-out block was excluded from the Stage-2 likelihood. The ordered GP, including covariance parameters, threshold-specific discrepancy variances, and spatially varying softplus gap surfaces, was estimated using only the exterior pseudo-observations. This experiment therefore evaluates recovery of an unobserved spatial block without oracle access to the generating gap surfaces.

Figure 7 shows one representative holdout realization. The ordered GP reconstructs the broad Moderate-exceedance pattern inside the outlined block, while posterior variance remains elevated within the unobserved area. The latent absolute-error map shows that the largest residual errors occur where direct pseudo-observations are absent, demonstrating appropriate spatial localization of uncertainty.

Across the 24 evaluations, the mean Brier score within withheld blocks was 0.0117, the mean log score was 0.5062, the mean threshold RMSE was 0.0943, and the mean latent-field RMSE was 0.3299. The mean posterior latent variance within the held-out blocks was 0.1324. Empirical coverage of nominal 90% latent-field intervals was 0.883 on average, with between-evaluation variability reflecting the limited number of buildings per block and differences in spatial connectivity among block locations. These results provide direct evidence that the fitted ordered model can recover an unobserved contiguous region while maintaining approximately calibrated uncertainty under model-consistent simulation.

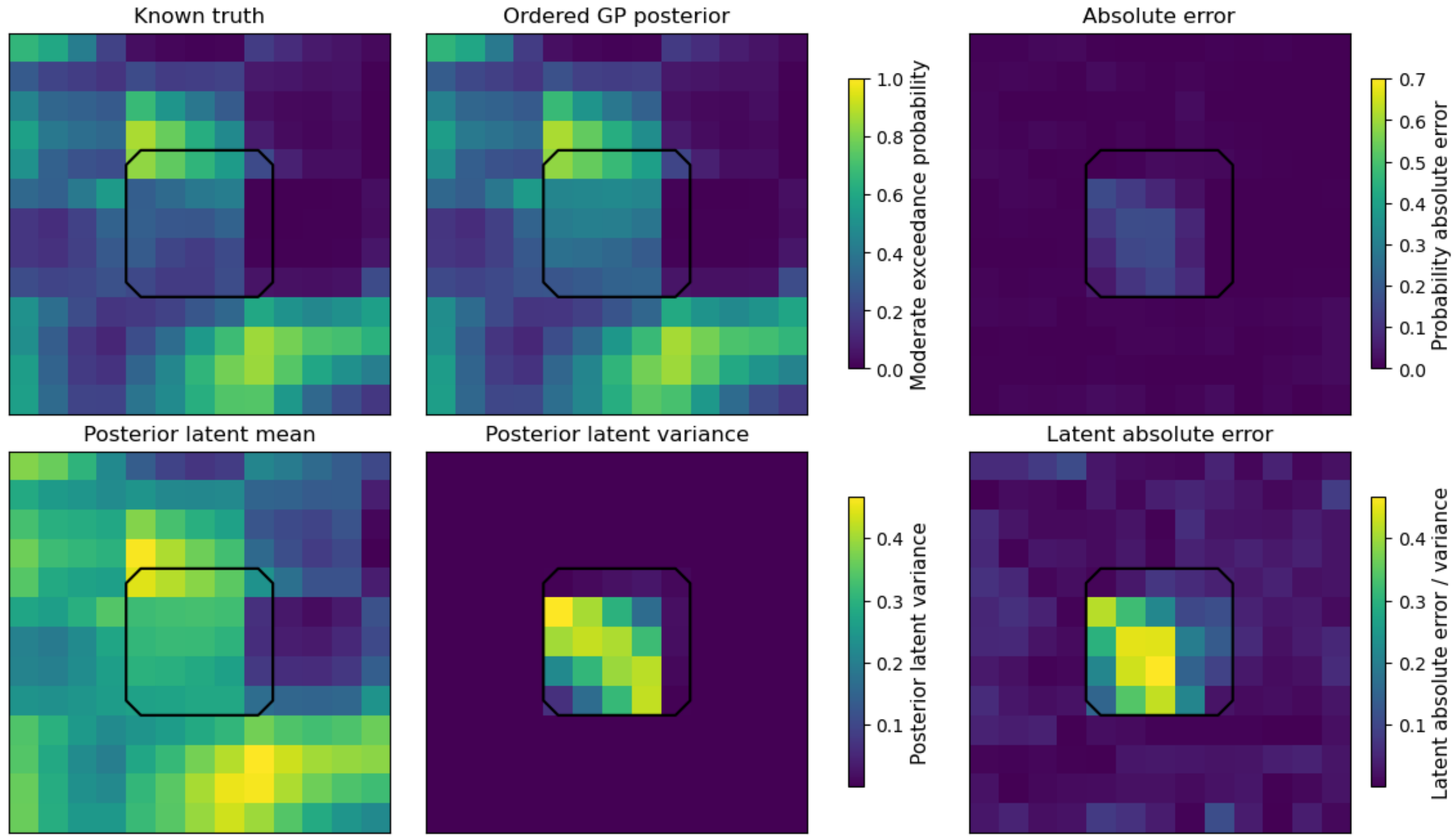

Figure 7: Representative model-consistent contiguous-block holdout. Top row: known Moderate-exceedance probability, ordered-GP posterior expected Moderate-exceedance probability, and absolute probability error. Bottom row: posterior mean of the shared latent field, posterior latent variance, and latent absolute error relative to the known latent field. The outlined block receives no direct pseudo-observations.

The fourth experiment evaluated the complete framework under a more demanding configuration. A biased PN prior was combined with noisy low-, medium-, and high-fidelity evidence, while the adjacent threshold-gap surfaces, covariance parameters, and discrepancy variances were estimated from the available data. The same predeclared contiguous blocks were withheld from direct evidence and excluded from the Stage-2 likelihood.

As summarized in Table 2, at evidence-supported buildings, the local PN–Beta–PN update reduced the Brier score from 0.0371 for the biased prior to 0.0177. The ordered heteroscedastic GP further reduced the Brier score to 0.0119. The threshold RMSE decreased from 0.1911 for the prior to 0.1326 after local updating and 0.1084 after ordered-GP propagation.

Within the information-gap region, the Stage-1 posterior remained equal to the biased prior because no local evidence was assimilated. The ordered GP nevertheless reduced the mean Brier score from 0.0385 to 0.0334 and the threshold RMSE from 0.1868 to 0.1717. These correspond to reductions of 13.2% and 8.1%, respectively. The smaller improvement relative to the model-consistent experiments is expected because this stress configuration combines prior misspecification, noisy heterogeneous evidence, estimated gap surfaces, and complete absence of direct block observations.

Table 2. Full end-to-end stress-test performance.

| **Method** | **Region** | **Brier** | **Brier SD** | **Log score** | **Log-score SD** | **Threshold RMSE** | **RMSE SD** |
|---|---|---|---|---|---|---|---|
| Biased PN prior | Information-gap region | 0.0385 | 0.0299 | 0.5157 | 0.1245 | 0.1868 | 0.0618 |
| Biased PN prior | Observed region | 0.0371 | 0.0091 | 0.5104 | 0.0482 | 0.1911 | 0.0239 |
| Local PN–Beta–PN | Observed region | 0.0177 | 0.0032 | 0.4620 | 0.0383 | 0.1326 | 0.0115 |
| Ordered heteroscedastic GP | Information-gap region | 0.0334 | 0.0307 | 0.5021 | 0.1208 | 0.1717 | 0.0636 |
| Ordered heteroscedastic GP | Observed region | 0.0119 | 0.0030 | 0.4440 | 0.0373 | 0.1084 | 0.0135 |
| Stage-1 posterior, no local evidence | Information-gap region | 0.0385 | 0.0299 | 0.5157 | 0.1245 | 0.1868 | 0.0618 |

The fifth experiment directly evaluated the ability of the ordered construction to prevent invalid threshold crossings when adjacent damage thresholds are difficult to distinguish. The ordered shared-field GP was compared with independently fitted threshold-wise GPs using the same withheld-block setting and sparse exterior evidence.

The ordered model produced no crossing posterior draws within the withheld block. The fitted mean adjacent gaps remained positive, with $\bar{\Delta}_{M\to E} = 0.5927$ and $\bar{\Delta}_{E\to C} = 0.3303$. In contrast, independently fitted threshold-wise GPs produced crossing latent-threshold realizations in 67.1% of withheld-block building-draw combinations. This diagnostic demonstrates that separate threshold-wise GP fits can produce invalid multistate posterior realizations under uncertainty, whereas the shared-field positive-gap construction preserves ordinality structurally.

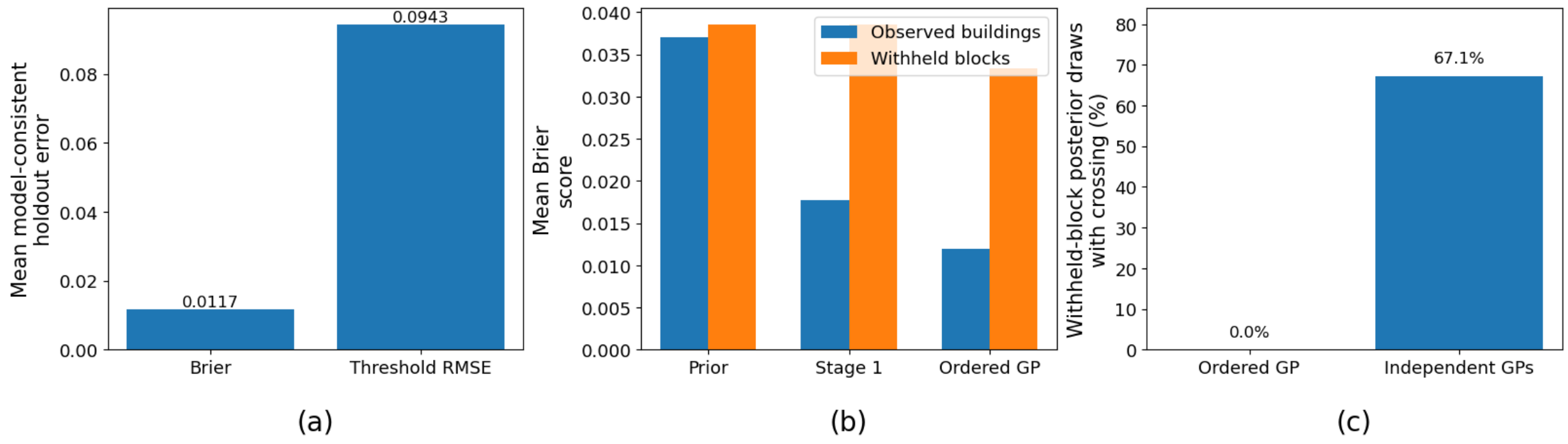

Figure 8 synthesizes the complementary validation results from Studies III–V. Panel (a) summarizes the repeated model-consistent block-holdout accuracy of the fully fitted ordered model in Study III. Panel (b) summarizes the observed-region and information-gap Brier scores from the more demanding end-to-end stress test in Study IV. Panel (c) reports the ordinal-crossing stress diagnostic from Study V, contrasting the ordered shared-field formulation with independently fitted threshold-wise GPs.

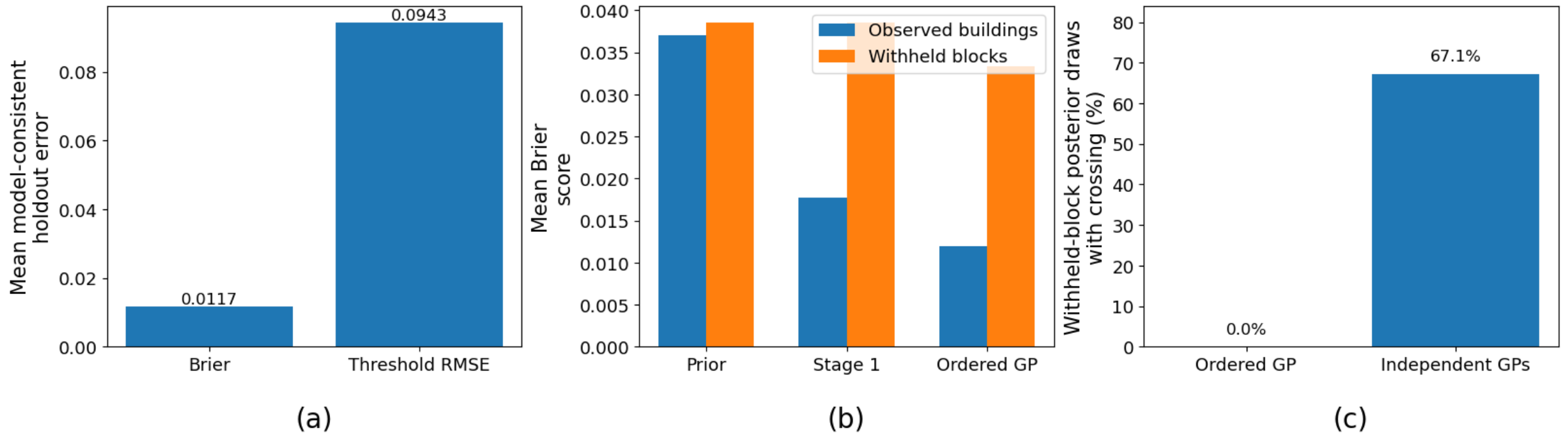

Figure 8: Cross-study summary of fitted-model recovery, end-to-end robustness, and ordinal validity. (a) Mean Brier score and threshold RMSE across the repeated model-consistent contiguous-block holdouts in Study III. (b) Mean Brier scores in observed buildings and withheld blocks under the full end-to-end stress test in Study IV. (c) Posterior-draw threshold-crossing rates within withheld blocks under the ordinal-crossing stress diagnostic in Study V.

## 4. Application Study

To demonstrate the applicability and scalability of the proposed methodology, the framework was applied to the Joplin testbed [78] for the 2011 Joplin tornado [79]. Figure 9(a) shows the spatial distribution of building archetypes, revealing a densely built inventory of 17,290 buildings dominated by wood-frame residential structures, with other archetypes distributed across the city. The archetypes are described in Table 3. Archetypes 1–4 were assigned the same parameter set because they correspond to small-to-medium wood-frame residential buildings represented by a common fragility classification in the source studies. Figure 9(b) overlays the historical tornado footprint and its Enhanced Fujita (EF) intensity contours. This spatially localized hazard setting provides a large-scale application context for examining the sequential behavior of the proposed updating framework. Figure 9(c) presents post-event NOAA remote-sensing imagery [80].

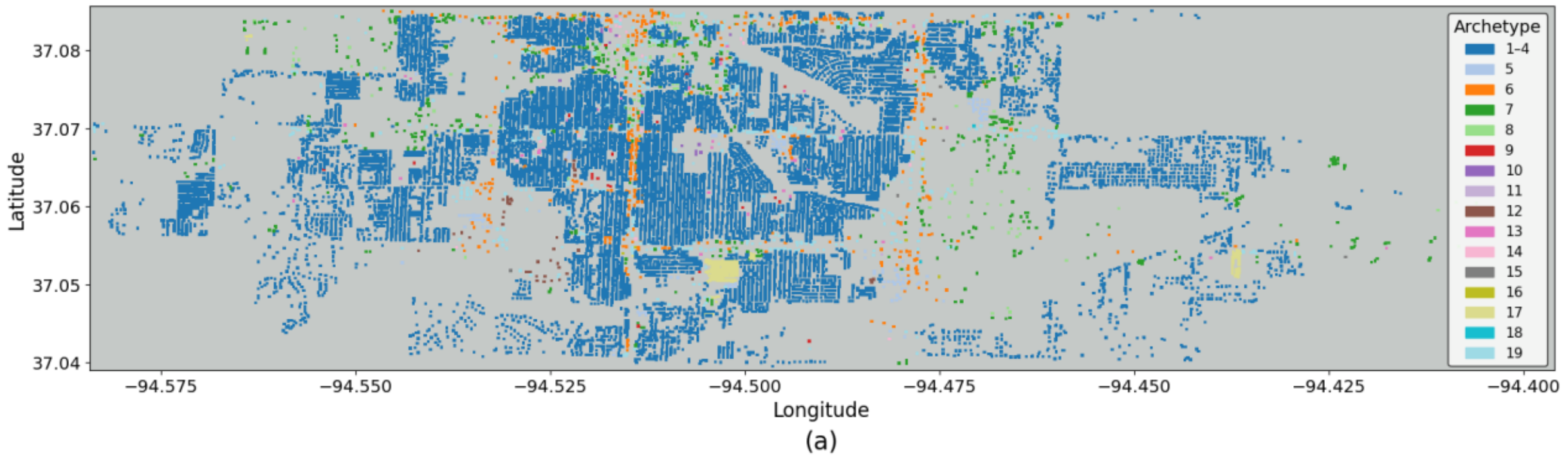

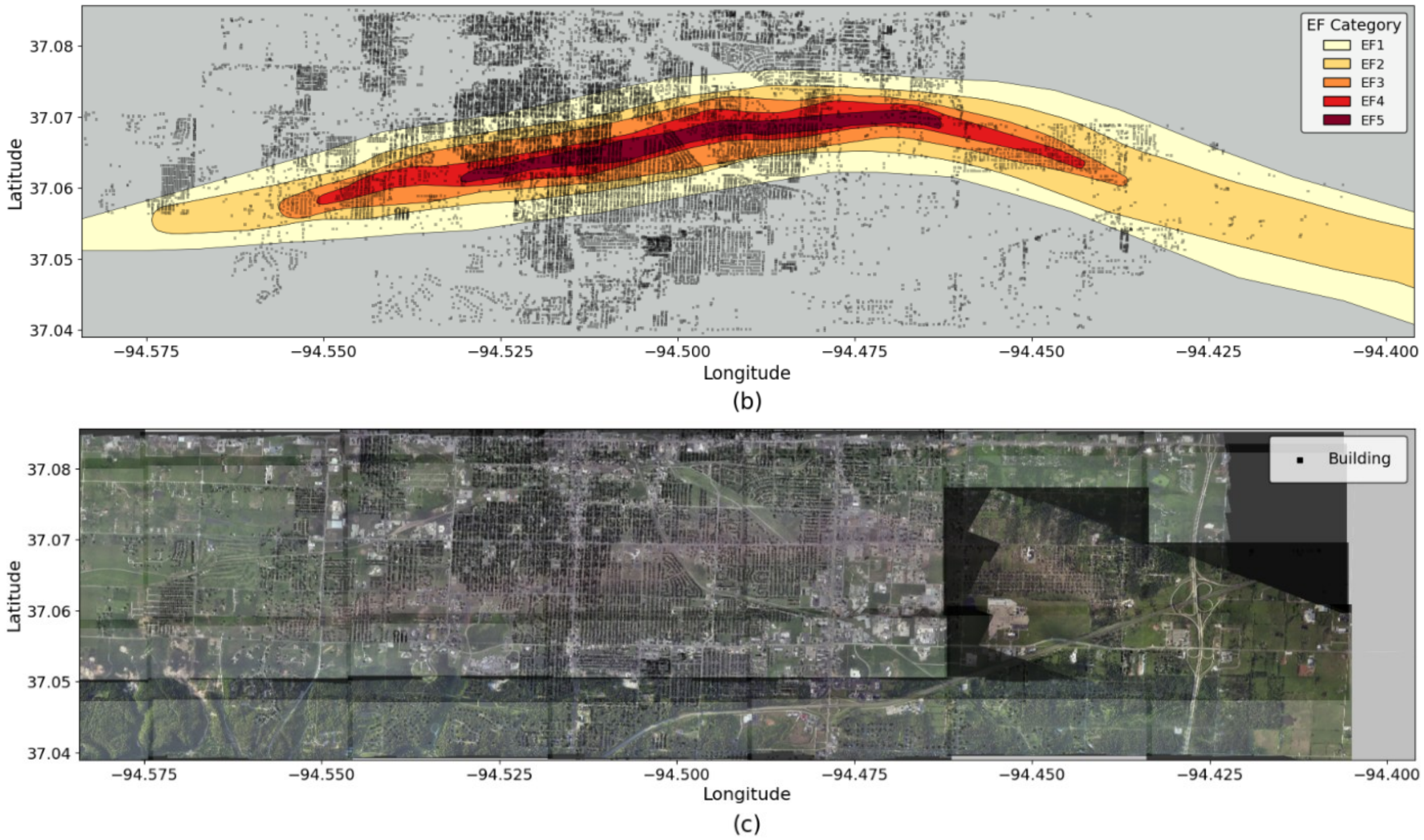

Figure 9: Maps showing (a) the spatial distribution of building archetypes, (b) EF intensity contours of the 2011 Joplin tornado obtained from IN-CORE [78], and (c) post-disaster remote-sensing imagery obtained from NOAA [80].

The resulting outputs are interpreted as event-conditioned posterior exceedance and exact-state probability fields, conditional on the adopted hazard prior, building inventory, physics-informed fragility assumptions, and available post-event evidence. They are not intended to represent hazard-independent fragility functions transferable without modification to other events. The remainder of this section is structured accordingly: §4.1 presents the development of the prior low-fidelity PN fragilities; §4.2 summarizes the computer-vision-based data used for updating; and §4.3 describes the experimental design and case-study comparisons.

### *4.1. Tornado Wind-Field Modeling for Prior Physics-Based Hazard Analysis*

Prior hazard characterization was developed using a simplified symmetric radial Rankine-type vortex model. It is assumed that, immediately after the event, the track centerline of the tornado can be identified from rapid-sensing products. Using this centerline, each building's perpendicular distance to the tornado axis is computed in projected (meter) coordinates. The tornado is represented as a radial wind field whose intensity depends solely on perpendicular distance from the identified track centerline.

A total tornado width $W_{tornado}$ is prescribed to represent the full lateral extent of damaging winds. In this study, $W_{\mathrm{tornado}}$ is treated as a scenario-dependent prior assumption rather than an observed constant tornado width. Multiple width scenarios are therefore used to represent uncertainty in the initial hazard characterization and to evaluate the sensitivity of subsequent updating to prior misspecification. For the EF5 prior scenario, the radial calibration was based on historical tornado-width proportions reported in [81]. In that study, the maximum-wind core and the radial location associated with the inner EF0 boundary correspond approximately to 27.3% and 87.3% of the total tornado width, respectively. Accordingly, for a prescribed width $W_{\mathrm{tornado}}$, the corresponding radii were defined as $R_{\mathrm{core}} = 0.273 W_{\mathrm{tornado}}/2$ and $R_{\mathrm{edge}} = 0.873 W_{\mathrm{tornado}}/2$, where $R_{\mathrm{core}}$ denotes the radius of the maximum-wind zone and $R_{\mathrm{edge}}$ denotes the radial location associated with the inner EF0 boundary. The associated calibration wind speeds were set to $V_{\mathrm{core}} = 115$m/s, representing a realistic upper-bound wind speed for an intense EF5 core, and $V_{\mathrm{edge}} = 38$m/s, corresponding to the EF0/EF1 transition wind speed.

Let $r$ denote the perpendicular distance from a site to the track centerline. The radial wind field is defined as:

$$V(r) = \begin{cases} V_{core} & , \quad r \leq R_{core} \\ V_{core}(R_{core}/r)^{\ln(V_{core}/V_{edge})/\ln(R_{edge}/R_{core})} & , \quad r > R_{core} \end{cases} \tag{53}$$

This power-law exponent enforces continuity between the EF5 core and the EF0 boundary, and the same decay is extended smoothly to the footprint boundary. Figure 10 shows the tornado wind field of various widths of 0, 400, 800, 1600, and 3200m.

The symmetric radial wind-field representation is intentionally simplified and does not capture tornado asymmetry, translation effects, storm-relative inflow, local turbulence, terrain effects, or along-track variation in vortex structure. It is therefore used only to construct a transparent physics-informed prior. The subsequent Bayesian updating refines the combined event-conditioned prior field using post-event evidence, but it does not separately identify whether disagreement arises from hazard-model error, fragility-model error, exposure uncertainty, or observational uncertainty.

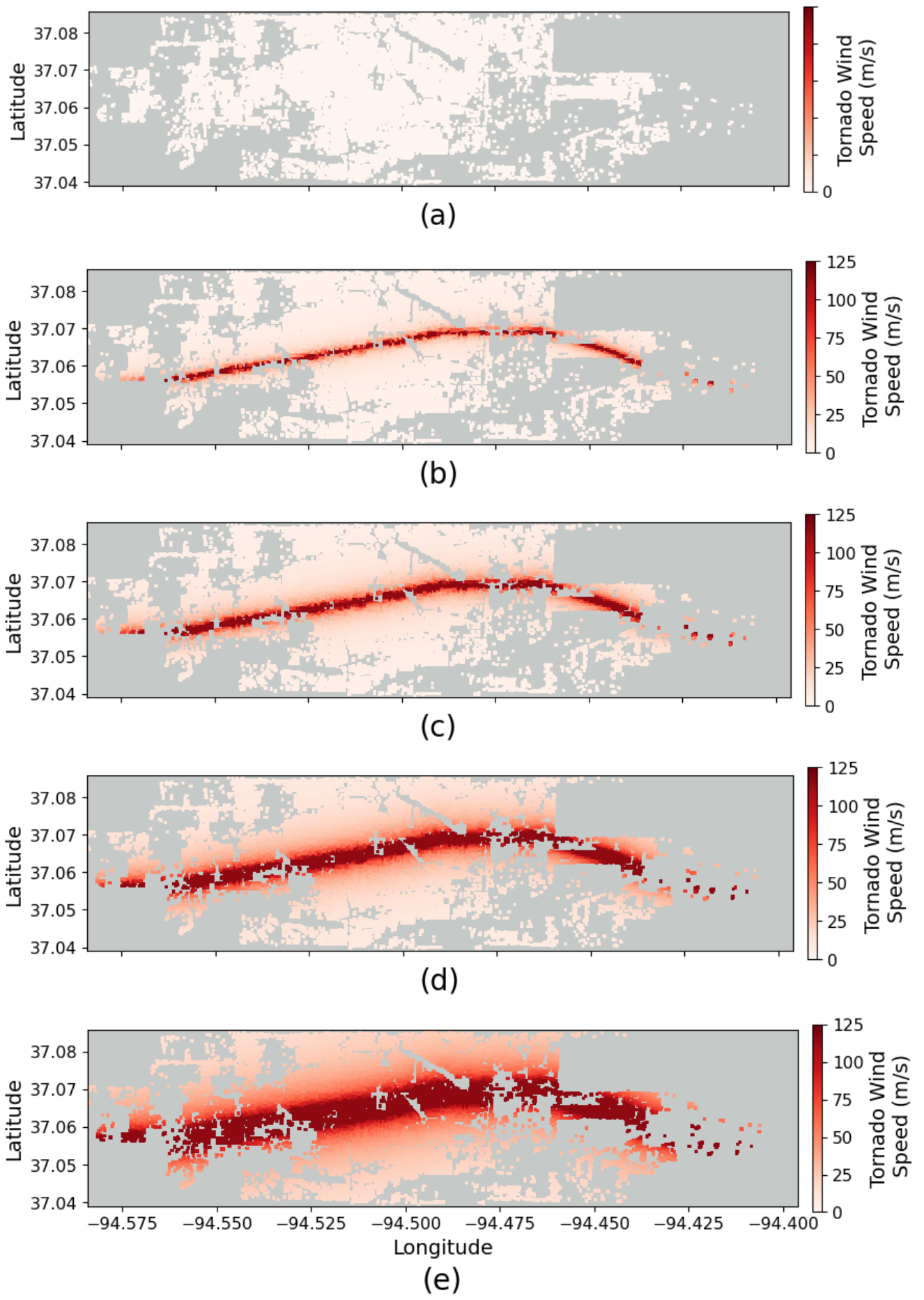

Figure 10: Maps showing the tornado wind field of various prior widths of (a) 0, (b) 400, (c) 800, (d) 1600, and (e) 3200m.

Using the constructed wind fields, lognormal fragility functions were applied to each building to compute damage-state exceedance probabilities and convert them into fragility indices in the probit space. The physics-based wind speeds were used as hazard-intensity inputs to the fragility functions. A log-scale hazard uncertainty of 0.09 was adopted based on the expert-based variability reported in the original EF documentation [82]. The fragility parameters, summarized in Table 3, were drawn from prior studies on tornado performance [23, 29-31]. Archetypes 1–4 were assigned the same parameter set because they correspond to small-to-medium wood-frame residential buildings represented by a common fragility classification in the source studies. A logarithmic capacity dispersion of 0.40 was applied across all archetypes and damage states [83].

Table 3: Tornado fragility model parameters for building archetypes and damage states [23, 29-31].

| Archetype | Damage State | Moderate | | Extensive | | Complete | |
|---|---|---|---|---|---|---|---|
| | Description | Median (m/s) | Dispersion | Median (m/s) | Dispersion | Median (m/s) | Dispersion |
| 1 - 4 | Small to medium residential wood building | 35.2 | 0.14 | 37.7 | 0.13 | 49.4 | 0.12 |
| 5 | Large residential wood building | 38.5 | 0.13 | 40.4 | 0.13 | 48.9 | 0.12 |
| 6 | Strip mall | 37.5 | 0.11 | 49.2 | 0.11 | 58.9 | 0.21 |
| 7 | Light industrial building | 40.2 | 0.1 | 44.0 | 0.1 | 47.7 | 0.1 |
| 8 | Heavy industrial building | 35.3 | 0.14 | 52.7 | 0.15 | 62.5 | 0.19 |
| 9 | Elementary/middle school | 42.3 | 0.11 | 49.2 | 0.1 | 65.0 | 0.12 |
| 10 | High school | 41.9 | 0.11 | 49.2 | 0.11 | 71.2 | 0.12 |
| 11 | Fire/police station | 49.2 | 0.12 | 57.1 | 0.12 | 65.7 | 0.19 |
| 12 | Hospital | 44.0 | 0.09 | 64.4 | 0.09 | 77.9 | 0.09 |
| 13 | Community center/church | 38.7 | 0.11 | 48.2 | 0.11 | 63.1 | 0.17 |
| 14 | Government building | 35.0 | 0.11 | 45.4 | 0.11 | 56.5 | 0.13 |
| 15 | Large big-box | 35.2 | 0.12 | 55.7 | 0.12 | 70.1 | 0.12 |
| 16 | Small big-box | 31.8 | 0.12 | 60.3 | 0.12 | 70.8 | 0.12 |
| 17 | Mobile home | 39.1 | 0.12 | 44.5 | 0.11 | 49.2 | 0.12 |
| 18 | Shopping center | 37.9 | 0.12 | 48.2 | 0.1 | 57.7 | 0.17 |
| 19 | Office building | 39.4 | 0.12 | 49.7 | 0.11 | 64.4 | 0.17 |

The resulting prior mean fragility indices quantify spatial vulnerability conditioned on the assumed tornado width. Figure 11 illustrates these prior fields for widths of 800 and 3200 m. For the 800m width (Figure 11a), large positive mean values, corresponding to high exceedance probability, are tightly concentrated near the track. For the 3200m scenario (Figure 11b), the high-fragility corridor becomes substantially broader. Ordinal exceedance probabilities are instead enforced structurally through the shared latent severity field and strictly positive threshold-gap functions described in §2.3.

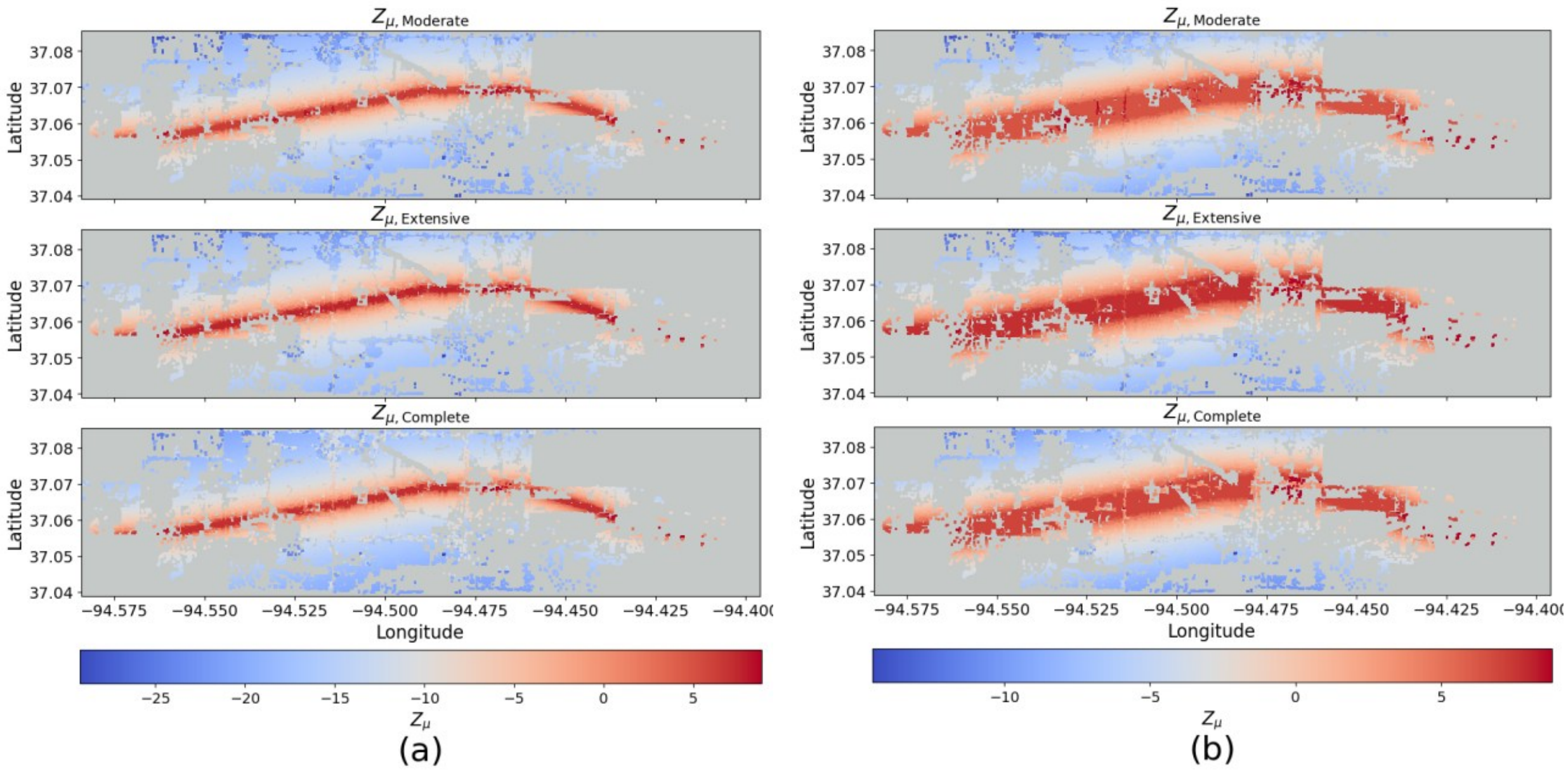

Figure 11: Spatial distribution of the prior physics-based mean fragility index in probit space for assumed tornado widths of (a) 800 m and (b) 3200 m.

Figure 12 presents corresponding expectations and variances in the probability space for the 800 m prior case. Expected exceedance probabilities follow the same spatial pattern as the probit means, but now lie within the [0,1] interval. Variance peaks near the transition zone between damaged and undamaged regions, where wind speeds lie near each damage-state threshold, and becomes small inside the core and far from the track.

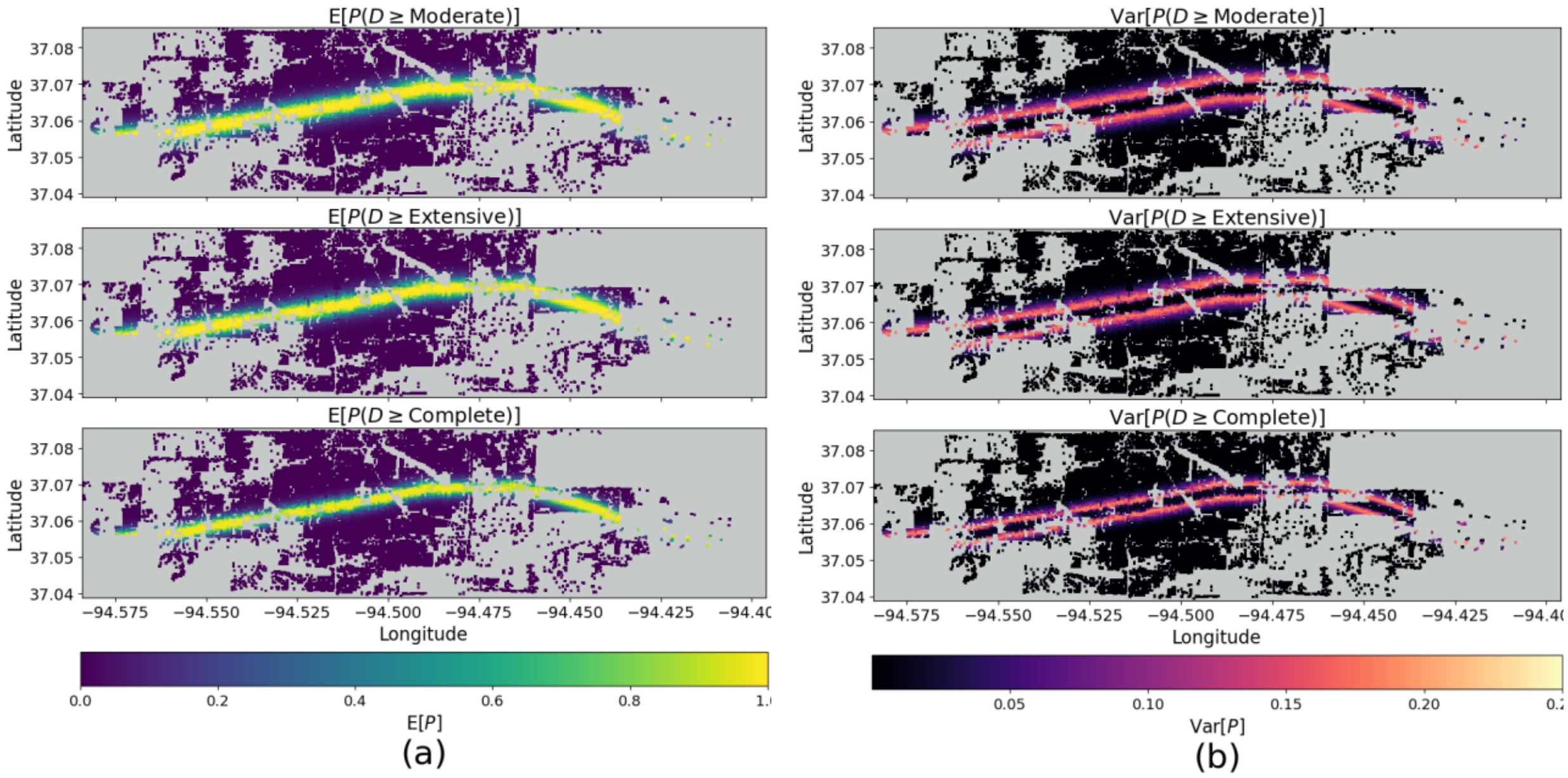

Figure 12: Spatial distribution of the prior fragility probabilities (transformed to probability space) for an assumed tornado width of 800 m: (a) expectation and (b) variance.

### *4.2. Updating with Computer-Vision–Based Damage Observations*

Damage classification was performed using a convolutional neural network (CNN) developed in a prior study by Braik and Koliou [61]. The model was trained using the xBD dataset [84] together with manually curated sub-images extracted from several NOAA post-disaster events (excluding Joplin). Its architecture consisted of sequential convolution–pooling blocks, followed by fully connected layers and a SoftMax output layer producing probabilistic predictions for the HAZUS-consistent damage states of Moderate, Extensive, and Complete. Training relied on cross-entropy minimization with standard augmentation and validation procedures, and the model's performance was evaluated using accuracy, precision, recall, F1-scores, and confusion matrices. In addition, the predictions were verified against historical post-Joplin damage observations. As the full development, training, and evaluation protocol has already been documented in Braik and Koliou [61], the reader is referred to that work for complete details.

The resulting soft confusion matrix and the exceedance-based confusion matrix are shown in Figure 13. The soft confusion matrix (Figure 13a) shows that the CNN reliably distinguishes between adjacent damage states, with strong diagonal values and limited confusion between low and high damage levels. When evaluated in terms of exceedance (Figure 13b), performance shows consistently high probabilities, reflecting the model's strength in identifying exceedance damage states.

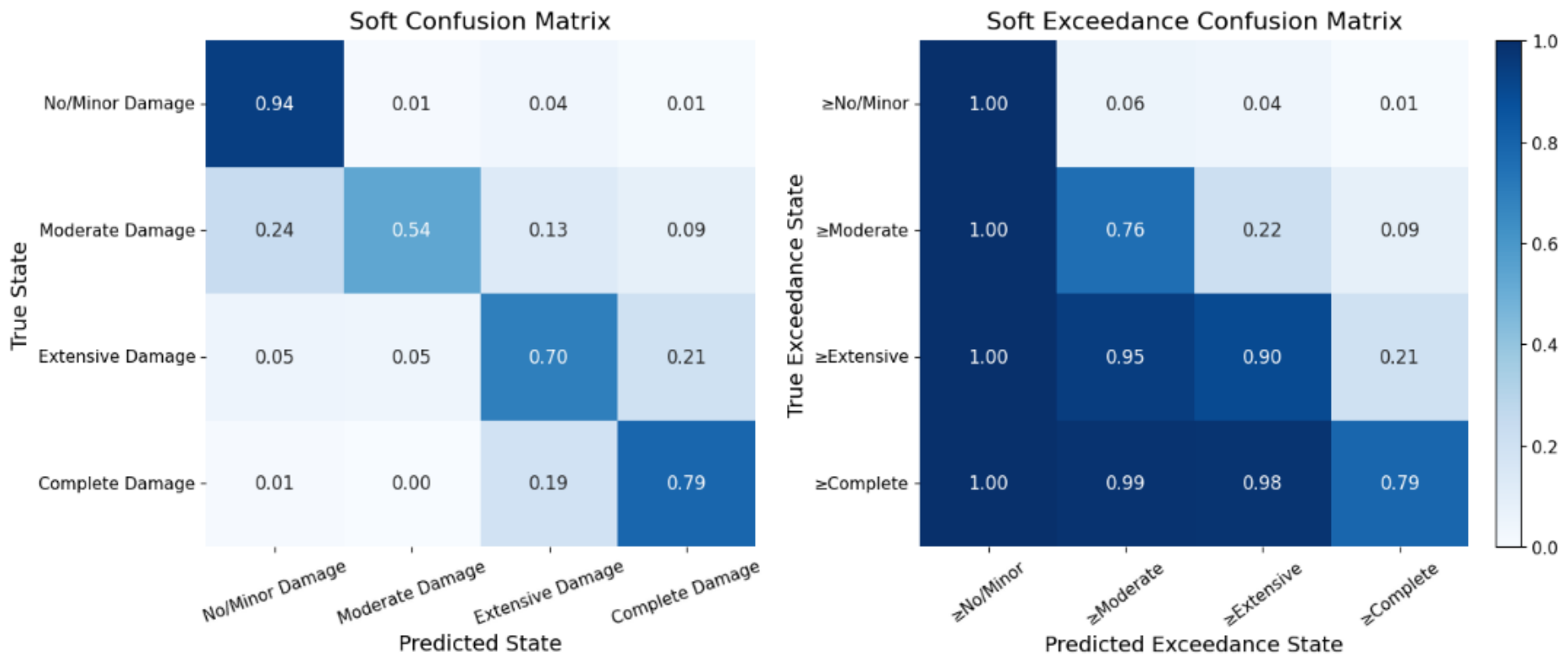

Figure 13: Soft confusion matrix and soft exceedance confusion matrix for the CNN-based tornado damage classifier.

The corresponding soft exceedance F1-scores and reliability weights (Table 4) confirm this pattern, with all thresholds providing meaningful probabilistic information for Bayesian updating. The resulting threshold-specific weights allow the CNN outputs to contribute as reliability-tempered soft evidence within the local Bayesian update.

The CNN outputs are treated as soft exceedance evidence rather than as ground-truth building-damage labels. Accordingly, the Joplin application evaluates how effectively the framework assimilates and spatially propagates the available computer-vision evidence; it does not provide an independent validation against complete building-level field damage labels. Although the evidence contributions differ in uncertainty and reliability, the Joplin application does not estimate an explicit stochastic discrepancy relationship between separate low- and high-fidelity simulators or observational systems. It is therefore more precisely interpreted as the sequential assimilation of heterogeneous evidence with differing reliability, rather than as classical multi-fidelity modeling.

Table 4: Soft exceedance F1-scores and information weights for each damage state.

| Damage State | Soft $F1_j$ | $w_j$ |
|---|---|---|
| Moderate | 0.90 | 6.68 |
| Extensive | 0.89 | 6.49 |
| Complete | 0.78 | 4.43 |

### *4.3. Experimental Design*

The experiment was designed to assess how the proposed sequential Bayesian framework responds to incrementally arriving computer-vision evidence. The full building inventory was randomly divided into an 80% sequential-assimilation subset and a 20% delayed-observation subset using a fixed random seed. All threshold observations associated with a building were assigned to the same subset to avoid cross-threshold leakage. Buildings in the sequential-assimilation subset received direct CNN evidence progressively through the intermediate update stages. The delayed-observation subset received no direct CNN evidence until the final update stage. Its pre-final trajectories therefore evaluate the extent to which the GP propagates information to buildings without local evidence before their own pseudo-observations become available. Because this split is random, it should not be interpreted as a test of performance across contiguous spatial observation gaps; instead, controlled spatial-block holdout validation is provided through the synthetic experiments in §3. Within the sequential-assimilation subset, CNN soft predictions were released in eight batches under two sampling strategies. Random sampling produced spatially scattered updates that mimic distributed reconnaissance. In contrast, grouped sampling used spatial K-means clustering with constrained cluster sizes to create geographically coherent batches.

To examine how the strength of the physical prior influences the updating process, prior fragility predictions were initialized using five possible tornado wind-field widths: 0, 400, 800, 1600, and 3200 m. At each batch, the Bayesian update was performed in two modes. In the first mode, only the buildings in the incoming batch were updated, reflecting a strictly local assimilation strategy. In the second mode, the GP was used to reshape the latent fragility field so that new evidence propagated spatially and structurally to all buildings, including those that had not yet received direct CNN predictions.

Sparse variational inference was implemented using 1,000 inducing points. The ordered GP parameters, including anisotropic spatial length scales, shared-field variance, archetype-correlation parameter, threshold-gap coefficients, threshold-specific discrepancy variances, and variational parameters, were optimized jointly using the Adam optimizer with a learning rate of $2 \times 10^{-2}$, minibatches of 4,096 pseudo-observations, and 250 optimization steps. Random seeds were fixed at 42 for reproducibility. Spatially grouped evidence batches were generated through K-means clustering using 10 initializations and the same fixed seed.

Model performance after each batch was evaluated using log score across the Moderate, Extensive, and Complete exceedance thresholds. Because the reference quantities are CNN-derived soft exceedance probabilities rather than verified field damage labels, this score measures agreement with the available computer-vision evidence under the adopted soft-label evaluation framework. It should not be interpreted as direct accuracy against true building damage states. Lower values indicate closer agreement with the CNN-derived probabilistic evidence.

### *4.4. Results and Discussion*

#### *4.4.1. Predictive Accuracy and Convergence*

Figure 14 shows the log-loss trajectories for both observed and unobserved buildings across the three damage states. Initial loss values reflect the assumed prior wind-field width: the 800 m and 1600 m priors start closest to the CNN evidence, followed by 400 m, while 0 m and 3200 m begin with noticeably higher error. This ordering aligns with the average EF5 width of roughly 837 m [85] and confirms that priors closer to this scale yield more calibrated initial predictions.

For the observed buildings, log-loss decreases steadily as batches of soft evidence are assimilated. The local Bayesian update reliably improves fragility estimates, and the sampling strategy has little effect on these curves, since improvements occur almost entirely at locations receiving direct observations.

The unobserved buildings provide a clearer test of GP propagation. Because they receive no direct evidence until the final batch, their early behavior depends entirely on how well the GP transfers information from the observed subset. Across all prior widths and sampling strategies, GP-enabled runs show consistent loss reduction despite the absence of local updates, indicating that the spatial–archetype kernel effectively diffuses local corrections into unobserved regions.

Sampling strategy matters more strongly for the unobserved subset. Under random sampling, evidence appears throughout the domain during each batch, allowing the GP to deliver smooth, widespread improvements. Under grouped sampling, updates are confined to a single region at a time, causing unobserved areas far from the updated clusters to improve more slowly until nearby evidence becomes available.

After the dashed vertical line, when unobserved buildings finally receive their first direct CNN predictions, the contrast between methods is clear. Local (non-GP) runs experience a sharp drop in loss because these buildings have remained identical to the prior until that moment. In GP runs, the curves change very little, demonstrating that most corrections had already been inferred and propagated before any local evidence arrived.

Across all priors, sampling strategies, and damage states, both observed and unobserved buildings converge to essentially identical final log-loss values once all evidence is assimilated. This convergence shows that prior hazard width and sampling order influence only the early data-scarce regime; with sufficient observations, the posterior is dominated by the available evidence and the GP's structured spatial–archetype correlations.

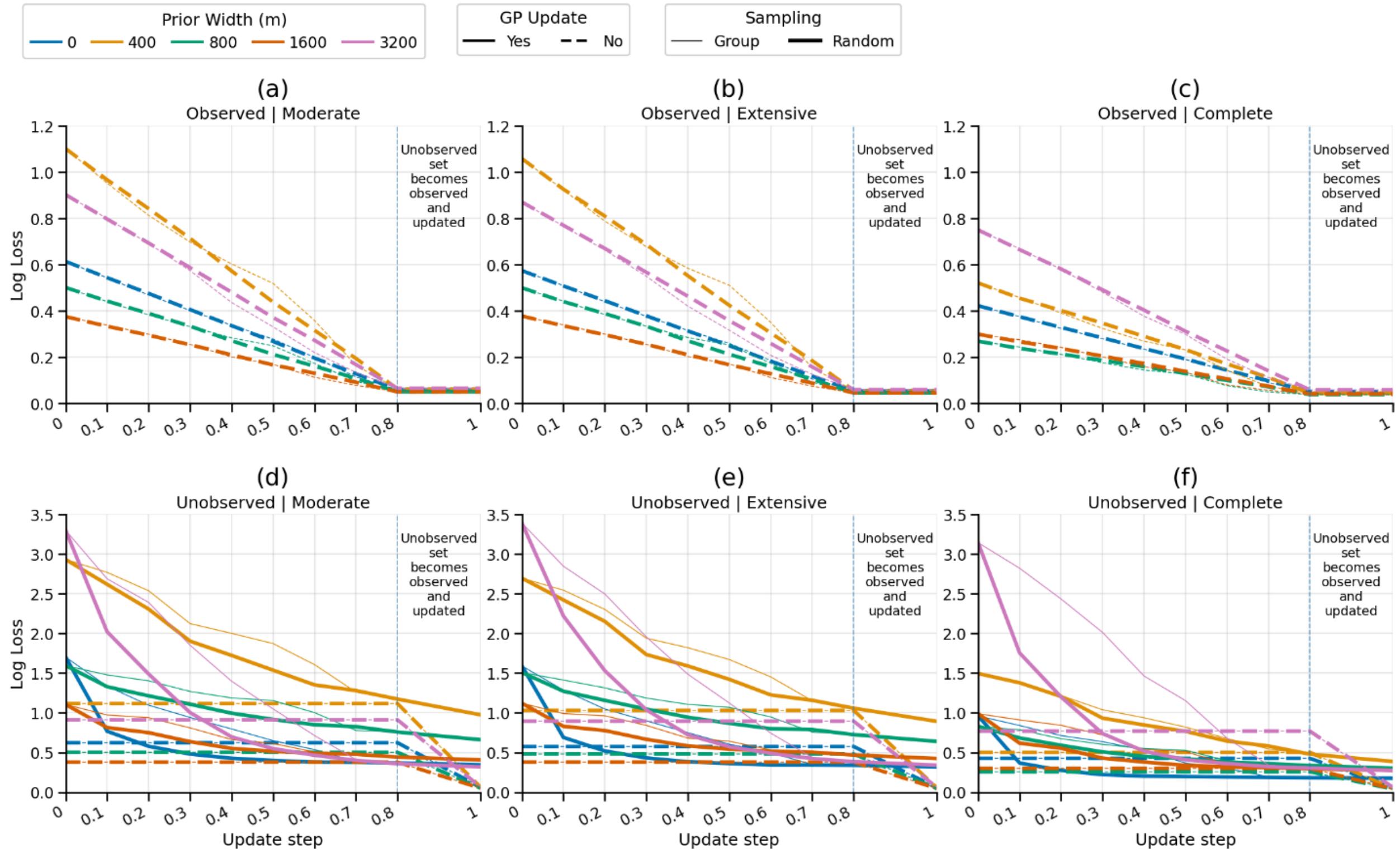

Figure 14: Log-loss evolution across update steps for observed (panels a–c) and unobserved (panels d–f) buildings. Columns correspond to the Moderate, Extensive, and Complete damage states, respectively. Results are shown under varying prior widths, sampling strategies, and with and without GP propagation.

### *4.4.2. Spatial Reconstruction and Uncertainty Quantification*

Figure 15 illustrates how the GP reconstructs the fragility field for unobserved buildings as CNN evidence is incrementally assimilated for the Extensive damage state, under the 0-m prior width, where the prior contains no spatial structure. At 0% updates, both sampling strategies produce a nearly uniform zero-probability field, reflecting the absence of a physics-based gradient. Under random sampling (Figure 15a), spatially dispersed evidence allows the GP to rapidly introduce high-fidelity corrections throughout the domain; the tornado corridor becomes visible as early as 10–20% updates, and by 40%, the field closely resembles the observed damage pattern due to efficient propagation across space and archetypes. Under grouped sampling (Figure 15b), early updates remain confined to the initial spatial cluster, producing strong local corrections while leaving the rest of the building portfolio unchanged at 10–20%. Only after later batches (80%) do these localized updates expand and merge into a coherent damage corridor. The 100% update corresponds to the moment when the previously unobserved buildings finally receive their own CNN-based pseudo-observations. As observed earlier in Figure 14, the GP has moved the delayed-observation buildings toward their final evidence-informed values before direct local evidence becomes available; the direct observations therefore produce only minor adjustments, and the two sampling strategies arrive at nearly identical spatial fields once the full data set is assimilated.

Figure 16 illustrates how the posterior exceedance probabilities and their associated uncertainties evolve for the Extensive damage state under an exaggerated prior tornado width of 3200 m, using random sampling. Panel (a) presents the expected exceedance probability for the unobserved buildings only, while panel (b) displays the heteroscedastic variance for all buildings, allowing the behavior of high- and low-fidelity subsets to be distinguished.

Because the 3200 m prior assumes an unrealistically wide tornado, the initial 0% field in Figure 16a shows high exceedance probabilities almost everywhere. As CNN evidence is incrementally assimilated (10–20%), the GP rapidly corrects these inflated predictions outside the damage corridor suggested by the CNN-derived soft evidence and reconstructs a much narrower, data-consistent pattern. By 40%, the posterior field is shaped almost entirely by the PN–Beta updates and GP propagation, substantially reducing disagreement with the CNN-derived soft evidence and producing a narrower evidence-consistent spatial gradient.

The corresponding variance fields in Figure 16(b) show how this correction develops. Initially, the variance is large throughout the region, reflecting the method's acknowledgment that the prior provides very little reliable information. As updates accumulate, variance drops sharply around buildings that have received CNN-based pseudo-observations and remains high elsewhere. By 80%, the map clearly separates regions that have been updated from those that have not.

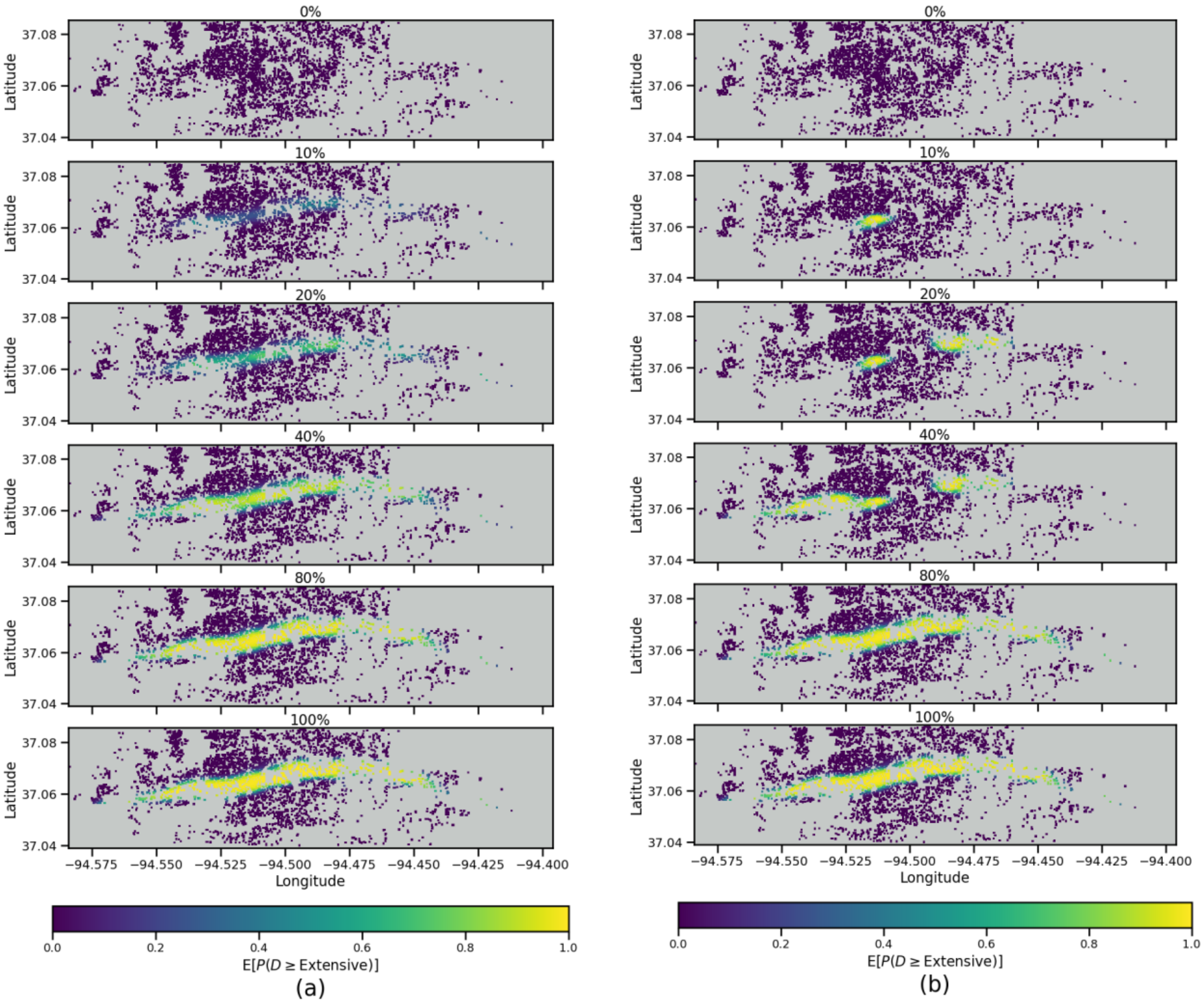

Figure 15: Incremental GP-based reconstruction of the Extensive-damage fragility field for unobserved buildings under the 0-m prior width, shown for random sampling (a) and grouped sampling (b).

At 100%, all buildings have been updated, and the remaining variance pattern is driven by both archetype and spatial position. An interesting feature appears along the edges of the reconstructed damage corridor: variance becomes higher near the boundaries. This occurs because the exceedance probabilities in these transition zones sit near the steepest part of the probit link, where small changes in the latent fragility index lead to large differences in probability.

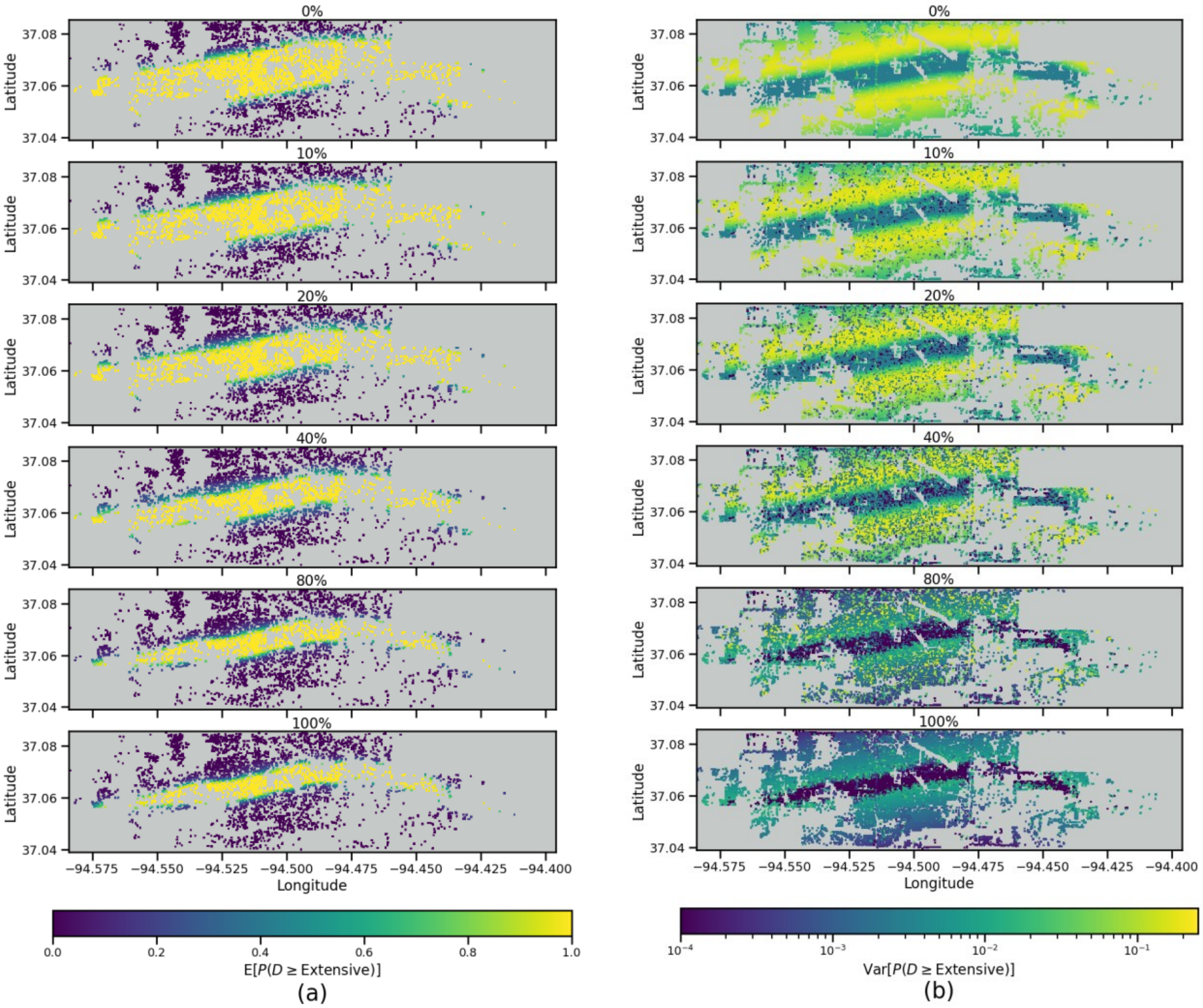

Figure 16: Evolution of the posterior exceedance field for the Extensive damage state under an exaggerated 3200 m prior tornado width with random sampling. Panel (a) shows the expected exceedance probability for the unseen buildings only, and panel (b) shows the corresponding variance for all buildings, at cumulative update levels of 0%, 10%, 20%, 40%, 80%, and 100%.

Figure 17 shows the distribution of posterior variance across all three damage states at the 80 percent update level, when 80 percent of buildings have already received CNN-based pseudo-observations and the remaining 20 percent are still unobserved. Figure 17(b) corresponds directly to the 80% variance map shown in Figure 16(b). Across all states, the variance distributions clearly separate the observed and unobserved buildings. Observed buildings show substantially smaller variance. This reflects the fact that these buildings have transitioned to high-fidelity estimates that strongly constrain the GP posterior. Unobserved buildings retain much larger variance because their predictions depend entirely on GP propagation from neighboring sites.

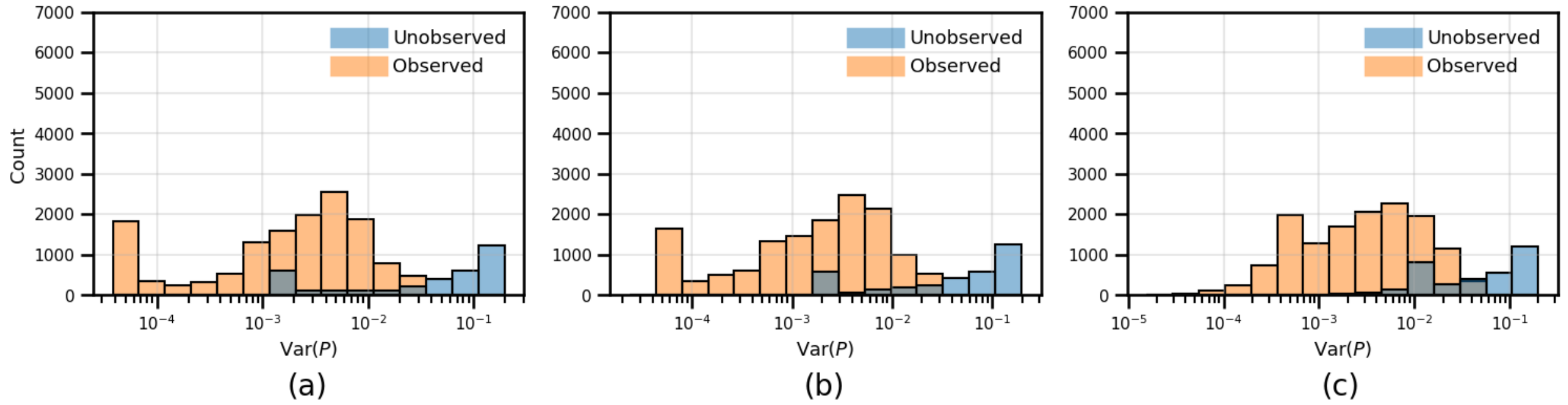

Figure 17: Posterior variance (Var($P$)) distributions for observed versus unobserved buildings for the (a) Moderate, (b) Extensive, and (c) Complete damage states.

### *4.4.3. Hyperparameter Evolution and Kernel Interpretation*

Figure 18 summarizes the evolution of the ordered-GP hyperparameters as successive batches of pseudo-observations are assimilated under different prior wind-field widths and evidence-arrival strategies. The spatial length scales, $\ell_{\text{lat}}$and $\ell_{\text{lon}}$, exhibit substantial adjustment during the earliest updates and then stabilize as additional evidence becomes available as shown in Figure 18(a) and Figure 18(b). The latitude length scale decreases rapidly from its initial values and converges to a relatively narrow range across scenarios. The longitude length scale follows a similar pattern after an initially wider spread, particularly for the widest prior scenarios. These trajectories indicate that the spatial correlation range is learned primarily during the early stages of assimilation, after which further updates refine rather than fundamentally alter the inferred spatial dependence.

The cross-archetype correlation parameter, $\rho_a$, remains consistently high across prior-width and sampling scenarios (Figure 18c). This suggests that the fitted model supports substantial information sharing among archetypes, while retaining $\rho_a < 1$ and therefore allowing archetype-specific differences to remain. This parameter should be interpreted as statistical dependence within the fitted spatial field rather than as a direct physical measure of similarity among structural systems.

The shared latent-field variance, $\sigma_F^2$, exhibits more visible dependence on the assumed prior wind-field width (Figure 18d). In particular, wider prior scenarios generally retain larger fitted field variance throughout the update sequence. This is expected because the width assumption changes the initial spatial scale and contrast of the physics-informed prior. Nevertheless, the trajectories remain bounded and progressively stabilize as evidence accumulates.

The threshold-specific discrepancy variances, $\tau_M^2$, $\tau_E^2$, and $\tau_C^2$, shown in Figure 18e–g, quantify residual variation between the local PN pseudo-observations and the ordered shared-field representation. Their evolution differs by threshold and prior scenario, indicating that the degree of disagreement between the local evidence and the global spatial model is not uniform across damage thresholds. In particular, wider prior scenarios tend to require larger discrepancy variances during early updates, consistent with greater mismatch between the corresponding prior fields and the assimilated pseudo-observations. These variances should not be interpreted as direct physical uncertainty in damage; rather, they represent model-scale residual uncertainty required to reconcile the local and global stages.

Figure 18(h) and Figure 18(i) show the mean fitted adjacent threshold gaps, $\bar{\Delta}_{M\to E}$ and $\bar{\Delta}_{E\to C}$, respectively. Both gaps remain positive throughout the sequential updates, as required by the structural ordering constraint. Their gradual evolution reflects adjustment of the separation between adjacent exceedance thresholds as the local pseudo-observations are assimilated. Figure 18(j) further shows that the minimum fitted adjacent gap across all buildings remains positive for every scenario and update stage. Therefore, ordinal exceedance probabilities are preserved by construction throughout the full sequential analysis, without clipping, manually imposed separation, or post hoc repair.

Overall, the hyperparameter trajectories show that the ordered GP adapts its spatial correlation structure, archetype dependence, discrepancy variances, and threshold-gap surfaces as evidence accumulates. The convergence patterns should not be interpreted as proof that the hyperparameters recover unique physical quantities; rather, they provide a diagnostic that the fitted covariance and ordering components remain stable, bounded, and internally consistent during sequential assimilation.

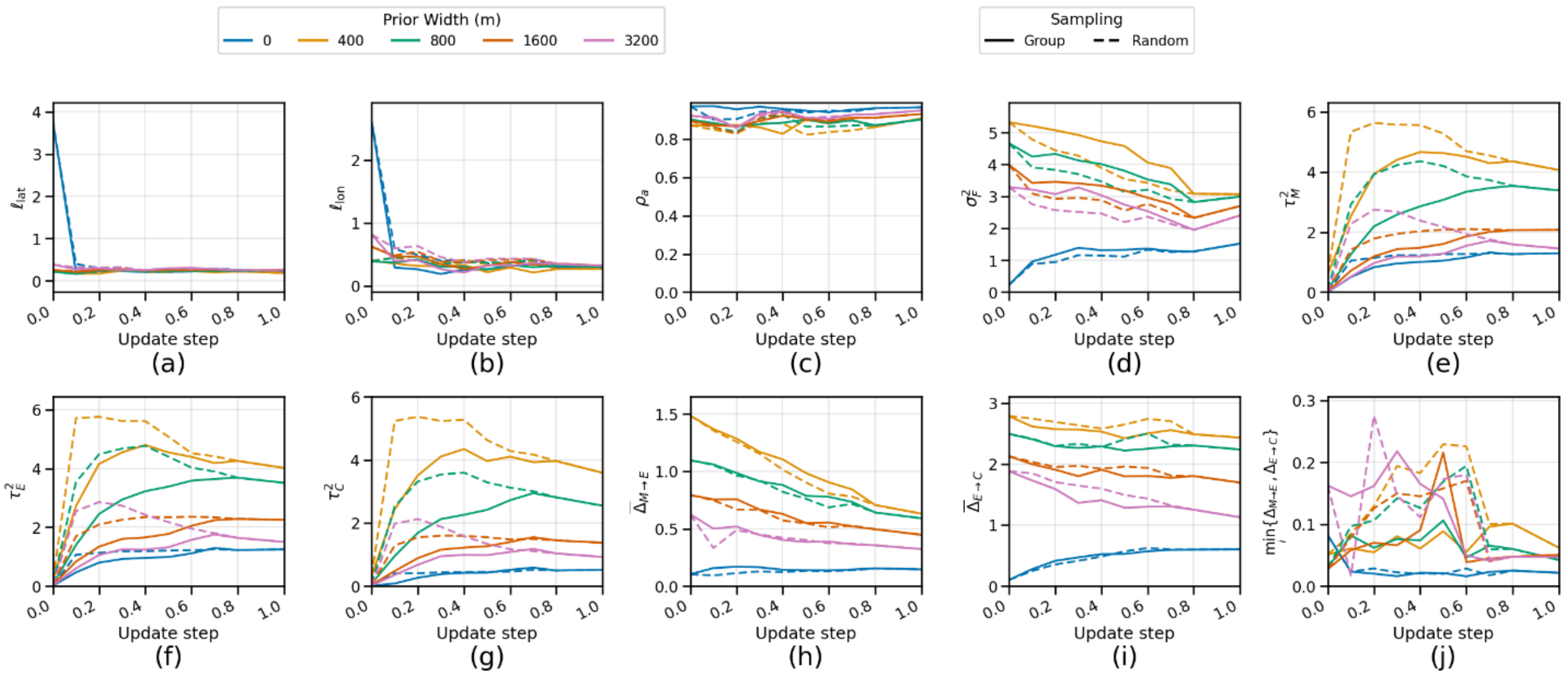

Figure 18: Evolution of ordered-GP hyperparameters across incremental updates for varying prior wind-field widths and evidence-arrival strategies: (a) latitude length scale, $\ell_{\text{lat}}$; (b) longitude length scale, $\ell_{\text{lon}}$; (c) cross-archetype correlation, $\rho_a$; (d) shared latent-field variance, $\sigma_F^2$; (e) Moderate-threshold discrepancy variance, $\tau_M^2$; (f) Extensive-threshold discrepancy variance, $\tau_E^2$; (g) Complete-threshold discrepancy variance, $\tau_C^2$; (h) mean positive threshold gap from Moderate to Extensive, $\bar{\Delta}_{M\to E}$; (i) mean positive threshold gap from Extensive to Complete, $\bar{\Delta}_{E\to C}$; and (j) minimum fitted adjacent threshold gap across buildings, $\min_i\{\Delta_{M\to E}(x_i), \Delta_{E\to C}(x_i)\}$. Solid and dashed lines denote grouped and random evidence-arrival strategies, respectively.

### *4.4.4. Computational Efficiency*

All analyses were conducted in Python on a desktop equipped with 64 GB of RAM and a 12th-generation Intel Core i7-12700 CPU operating at 2.1 GHz. Prior hazard and fragility construction, PN formulation, PN–Beta moment matching, conjugate Beta updating, and associated numerical routines were implemented using NumPy, SciPy, GeoPandas, and PyTorch. For the controlled synthetic study, the ordered GP was implemented directly in PyTorch and estimated through the exact Gaussian marginal likelihood. This exact implementation was computationally feasible because the synthetic inventory contained 169 buildings and three exceedance thresholds, corresponding to 507 stacked pseudo-observations. For the Joplin application, sparse variational GP inference was implemented using GPyTorch [86] to accommodate the larger regional inventory. For 17,290 buildings and three exceedance thresholds, each sequential updating step—including local PN–Beta–PN updating and global sparse variational ordered-GP optimization—required approximately 8 minutes under the computational configuration described in §4.3. These timings demonstrate that the proposed framework is computationally practical for sequential regional-scale updating, while the exact synthetic implementation was retained to permit controlled validation without sparse-GP approximation.

## 5. Conclusions and Future Work

This study developed a two-stage Bayesian framework for sequential updating of event-conditioned spatial fragility fields using physics-informed priors and post-disaster evidence of differing reliability. The framework links pre-event vulnerability modeling with incrementally available post-event observations while retaining uncertainty quantification, spatial dependence, and probabilistically valid multistate damage representations.

In the first stage, physics-informed fragility estimates are represented in Probit–Normal (PN) form, retaining both the central tendency and uncertainty of prior exceedance probabilities on a latent Gaussian scale. Each PN marginal is moment-matched to a Beta surrogate, updated using soft exceedance evidence through reliability-tempered power-likelihood contributions, and mapped back into PN form. This computationally tractable local update produces building- and threshold-specific posterior uncertainty that reflects the amount, quality, and consistency of available evidence. The resulting Stage-1 estimates provide heteroscedastic pseudo-observations for the spatial model.

In the second stage, the local pseudo-observations are assimilated through an ordered spatial GP. Rather than fitting separate unconstrained GP fields for each damage threshold, the proposed formulation uses a shared latent spatial severity field and strictly positive adjacent threshold-gap functions. This construction guarantees pathwise ordering of exceedance probabilities, conditional on the fitted gap surfaces, and therefore yields nonnegative exact damage-state probabilities without clipping latent values, manually imposing threshold separation, or applying post hoc ordinal repair. The spatial model enables information sharing across nearby buildings and structurally related archetypes while preserving elevated uncertainty where local evidence is limited.

The controlled synthetic experiments established complementary properties of the framework. First, under model-consistent spatial propagation, increasingly dense and spatially connected exterior evidence improved reconstruction within an unobserved contiguous region before any direct interior evidence was introduced. The mean gap-region Brier score decreased from 0.0591 with no evidence to 0.0127 after complete exterior coverage, while latent-field RMSE decreased from 0.9207 to 0.4320 and mean posterior variance decreased from 0.9000 to 0.1829. Limited interior anchors further reduced the Brier score and latent-field RMSE to 0.0018 and 0.1487, respectively.

Second, the local PN–Beta–PN update appropriately differentiated evidence sources through calibrated reliability weights. Low-fidelity evidence alone did not improve the biased prior in aggregate, but subsequent assimilation of medium- and high-fidelity evidence reduced the Brier score from 0.0224 to 0.0179 and the threshold RMSE from 0.1497 to 0.1337. This result demonstrates that reliability tempering limits the effect of weak evidence while allowing sufficiently informative evidence to improve local posterior estimates.

Third, repeated model-consistent contiguous-block holdouts demonstrated recovery by the fully fitted ordered GP without oracle access to the generating gap surfaces. Across 24 predeclared block evaluations, the mean withheld-block Brier score was 0.0117, threshold RMSE was 0.0943, and latent-field RMSE was 0.3299. Empirical coverage of nominal 90% latent-field intervals was 0.883 on average, indicating approximate calibration under the model-consistent setting.

Fourth, the full end-to-end stress test combined a biased PN prior, noisy heterogeneous evidence, estimated gap surfaces, and contiguous regions receiving no direct observations. At evidence-supported buildings, the Brier score decreased from 0.0371 for the biased prior to 0.0177 after local updating and to 0.0119 after ordered-GP propagation. Within the information-gap region, the ordered GP reduced the Brier score from 0.0385 to 0.0334 and threshold RMSE from 0.1868 to 0.1717, demonstrating covariance-supported improvement under a deliberately demanding configuration.

Finally, the ordinal-crossing stress diagnostic demonstrated the structural advantage of the shared-field positive-gap formulation. The ordered GP produced no threshold crossings across posterior draws in the withheld block, whereas independently fitted threshold-wise GPs produced crossing latent-threshold realizations in 67.1% of withheld-block building-draw combinations. Thus, the proposed formulation does not merely encourage ordinal behavior; it enforces valid multistate posterior realizations by construction.

The Joplin tornado application demonstrated the scalability and practical behavior of the framework for a regional inventory exceeding 17,000 buildings. Simplified radial wind-field scenarios were used to construct physics-informed priors, and CNN-derived soft damage predictions were assimilated as post-event evidence. The local update reduced disagreement with the available CNN-derived evidence at directly updated buildings, while the ordered spatial GP propagated locally updated information across the inventory. The resulting fields became less sensitive to the assumed initial tornado-width scenario as evidence accumulated. However, these outputs should be interpreted as event-conditioned posterior exceedance and exact-state probability fields conditional on the adopted wind-field prior, inventory assumptions, fragility formulation, and computer-vision evidence. They do not separately identify hazard-model error, fragility-model error, exposure uncertainty, or observation error, nor do they constitute independent validation against complete building-level field damage labels.

Several limitations and future directions remain. The present framework accommodates evidence sources of differing reliability through calibrated power-likelihood weights, but it does not estimate an explicit stochastic discrepancy relationship between low- and high-fidelity sources. Future work could introduce hierarchical or discrepancy-process formulations to learn source-specific bias and uncertainty directly from data. A fully Bayesian treatment of covariance parameters, threshold-gap coefficients, and discrepancy variances would also propagate hyperparameter uncertainty beyond the present optimized-parameter implementation.

The spatial model further assumes stationary anisotropic dependence and an exchangeable archetype-correlation structure. More flexible nonstationary kernels, locally varying correlation structures, and richer building descriptors may improve performance in settings with abrupt hazard gradients or substantial unobserved building-level heterogeneity. Future validation should incorporate geographically complete field damage labels, additional hazard events, and other observational sources, including inspections, crowdsourced imagery, sensor networks, and structural analyses.

Overall, the study provides a structured approach for evolving static physics-informed fragility priors into uncertainty-aware event-conditioned spatial posterior fields as post-disaster evidence becomes available. By combining reliability-tempered local Bayesian updating with structurally ordered spatial propagation, the framework provides a basis for regional damage assessment systems that are sequential, spatially coherent, uncertainty-aware, and probabilistically consistent across damage states.

**Declaration of competing interests**

The authors declare that they have no known competing financial interests or personal relationships that could have appeared to influence the work reported in this paper.

**Acknowledgments**

Partial financial support for this study was provided by the US National Science Foundation (NSF) under Award Number 2052930. This financial support is gratefully acknowledged. Any opinions, findings, conclusions, and recommendations presented in this paper are those of the authors and do not necessarily reflect the views of NSF.

**Appendix A: Numerical Validation of the PN–Beta–PN Local Posterior Approximation**

This appendix evaluates the accuracy and computational efficiency of the PN–Beta–PN local updating procedure introduced in §2.2.1. The approximation is assessed under the corresponding powered soft-evidence likelihood across supportive, contradictory, ambiguous, and conflicting multi-fidelity evidence regimes.

Let:

$$P = \Phi(Z), \qquad Z \sim \mathcal{N}(\mu_0, \sigma_0^2) \qquad \text{(A1)}$$

where $Z$ is the latent probit-scale fragility index and $P$ is the associated exceedance probability. For soft exceedance observations $\{y^{(b)}\}_{b=1}^{B}$ and reliability weights $\{w^{(b)}\}_{b=1}^{B}$, the exact local posterior on the latent scale is:

$$p(z \mid y, w) \propto \phi(z; \mu_0, \sigma_0^2) \prod_{b=1}^{B} \Phi(z)^{w^{(b)} y^{(b)}} [1 - \Phi(z)]^{w^{(b)}[1-y^{(b)}]} \quad \text{(A2)}$$

Because this posterior is one-dimensional, convergence-controlled numerical quadrature was adopted as the primary numerical reference. Posterior probability moments were computed by integrating the normalized posterior density after progressively tightening the numerical integration tolerance. A quadrature result was retained only after the posterior mean and variance of $P = \Phi(Z)$were stable under the refinement criterion used in the numerical implementation.

For the PN–Beta–PN approximation, the prior distribution of $P$is first moment-matched to $\mathrm{Beta}(\alpha, \gamma)$. The Beta parameters are then updated as:

$$\alpha' = \alpha + \sum_{b=1}^{B} w^{(b)} y^{(b)}, \qquad \gamma' = \gamma + \sum_{b=1}^{B} w^{(b)} \left[1 - y^{(b)}\right] \quad \text{(A3)}$$

The updated Beta posterior is subsequently reprojected into a Probit–Normal representation by matching its posterior mean and variance. Thus, the Beta update itself is analytic, whereas the final Beta-to-PN reprojection is a deterministic moment-matching calculation.

The numerical comparison was conducted over $\mu_0 \in [-3,3]$ and $\sigma_0^2 \in [0,3]$ under the four evidence regimes described above. For each configuration, posterior-mean error, posterior-variance error, and Kullback–Leibler (KL) divergence were evaluated relative to the converged quadrature posterior. Runtime was separately evaluated for PN–Beta–PN, converged quadrature, and converged MCMC.

Figure A. 1 presents the posterior-moment error of PN–Beta–PN relative to the converged quadrature solution. Across all evidence regimes, the approximation is most accurate in the transition region, where prior exceedance probabilities are not concentrated near zero or one. The largest posterior-mean and posterior-variance discrepancies occur in boundary-dominated

configurations involving small prior variance and strongly conflicting evidence. In these regimes, the posterior probability distribution can become highly concentrated near zero or one, making the exact probability-scale shape more sensitive to the PN–Beta moment-matching approximation.

Nevertheless, the approximation remains accurate over the principal operational region in which posterior uncertainty is substantial, and the fragility estimate is most informative for sequential updating. The supportive and contradictory evidence regimes exhibit the expected approximate mirror symmetry under sign reversal of the latent probit scale. The conflicting multi-fidelity regime produces localized discrepancies near boundary-dominated configurations but preserves small errors over the broad central region of the parameter space.

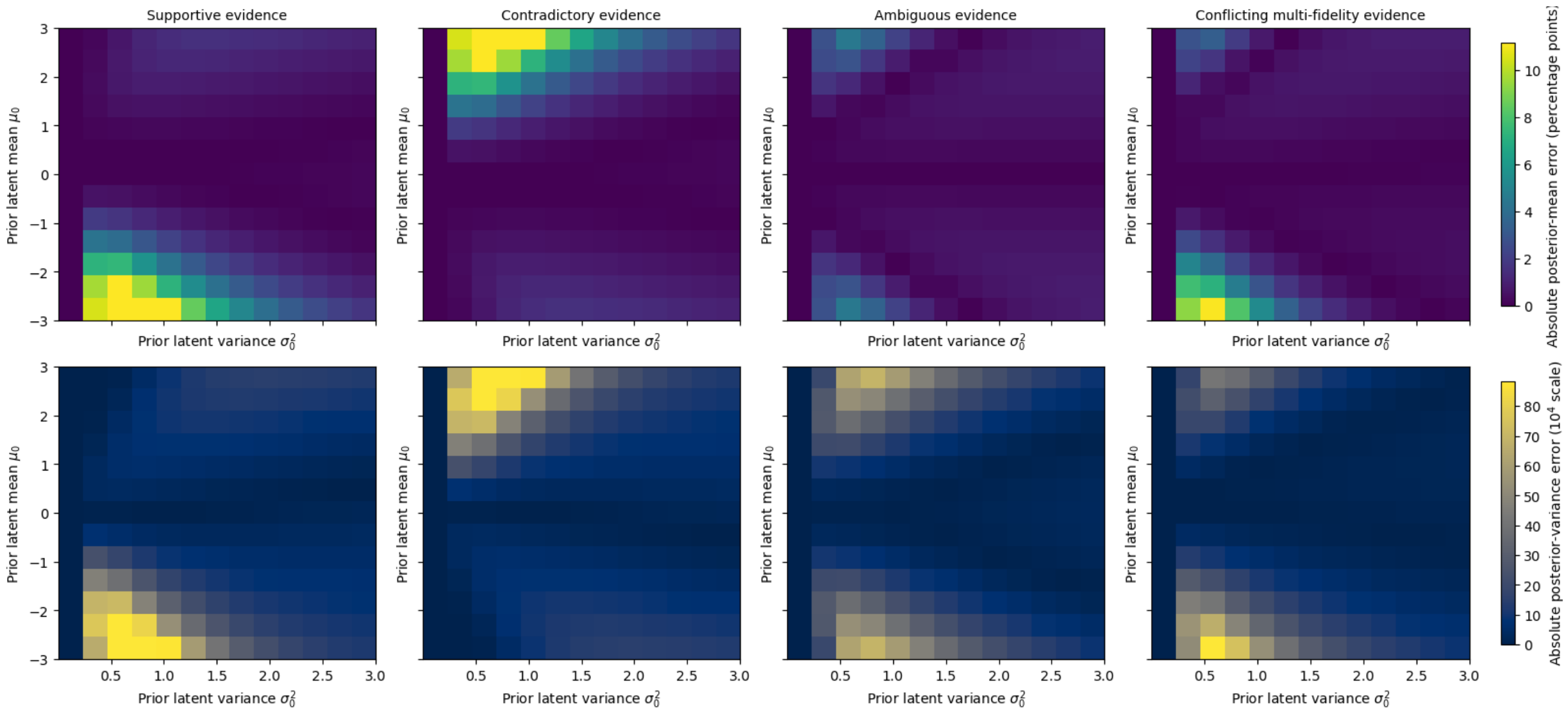

Figure A. 1. Absolute posterior-mean error and absolute posterior-variance error of the PN–Beta–PN approximation relative to the converged quadrature benchmark across prior latent means, prior latent variances, and evidence regimes. Posterior-mean error is expressed in percentage points. Posterior-variance error is displayed on a logarithmic scale.

Figure A. 2 compares the runtime of convergence-controlled numerical quadrature with that of PN–Beta–PN. Define the quadrature runtime ratio as:

$$R_{\text{quad}} = t_{\text{quad}} / t_{\text{PN-Beta-PN}} \quad \text{(A4)}$$

Across the full parameter and evidence grid, the 10th-percentile, median, and 90th-percentile values of $R_{\text{quad}}$ were approximately $49.1\times$, $51.6\times$, and $55.4\times$, respectively. Thus, under the adopted convergence-control criterion, the PN–Beta–PN procedure was consistently much faster than numerical quadrature while retaining close agreement with its posterior moments.

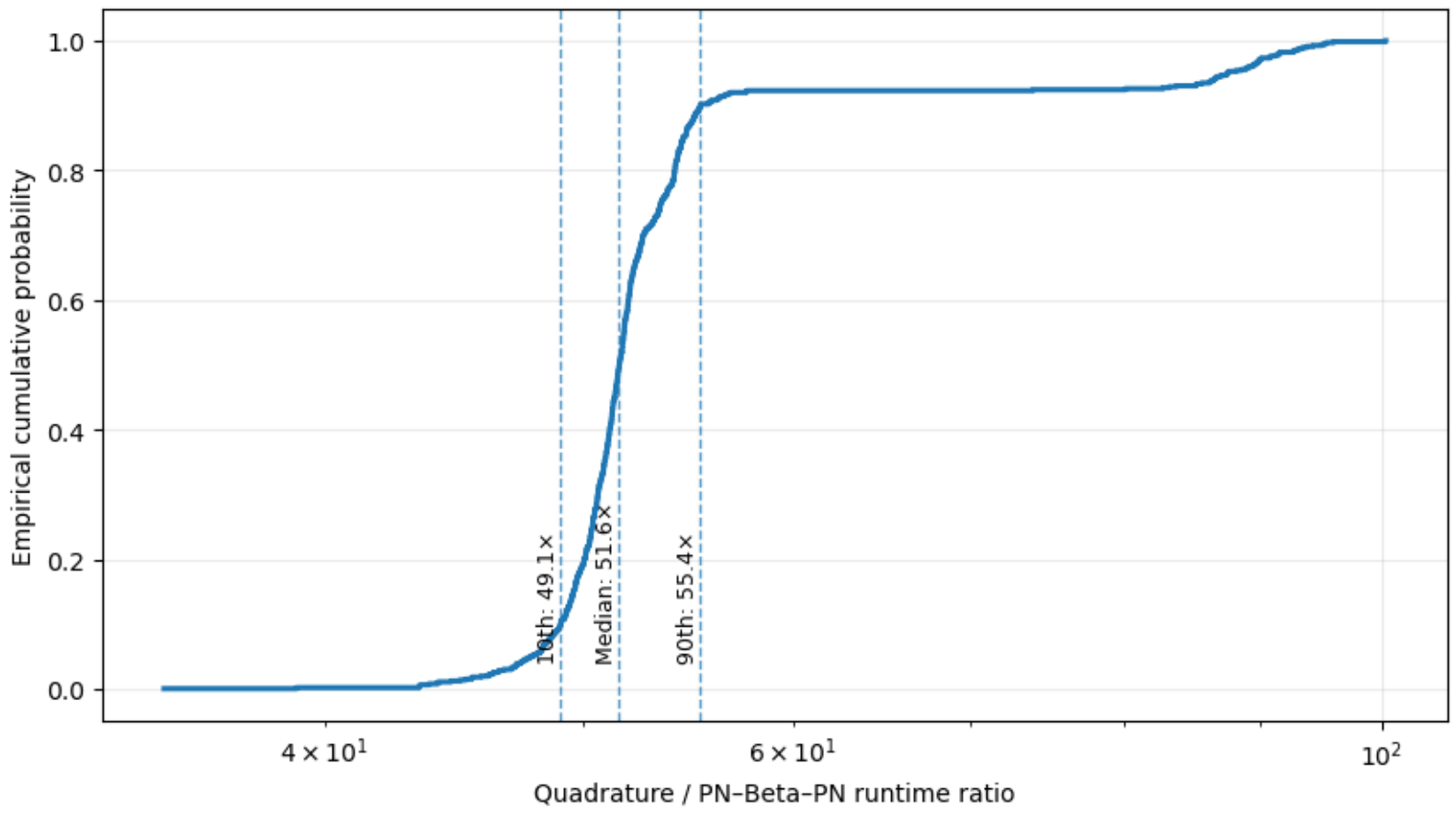

Figure A. 2. Empirical cumulative distribution of the converged-quadrature-to-PN–Beta–PN per-update runtime ratio over the full numerical experiment.

Although converged quadrature is the appropriate accuracy reference for the present scalar posterior, MCMC was additionally included as a generic sampling-based computational benchmark. Each MCMC result was retained only after satisfying the convergence and Monte Carlo precision criteria specified in the numerical implementation. MCMC is not used as the accuracy reference because, for this smooth one-dimensional posterior, quadrature provides a deterministic tolerance-controlled solution without sampling error.

Define the MCMC runtime ratio as:

$$R_{\mathrm{MCMC}} = t_{\mathrm{MCMC}}/t_{\text{PN-Beta-PN}} \qquad \text{(A5)}$$

Figure A. 3 shows the empirical distribution of this ratio across the common parameter and evidence grid. The 10th-percentile, median, and 90th-percentile MCMC-to-PN–Beta–PN runtime ratios were approximately $136.8\times$, $211.1\times$, and $293.4\times$, respectively. Therefore, MCMC required substantially greater computation than PN–Beta–PN even after restricting comparison to converged chains.

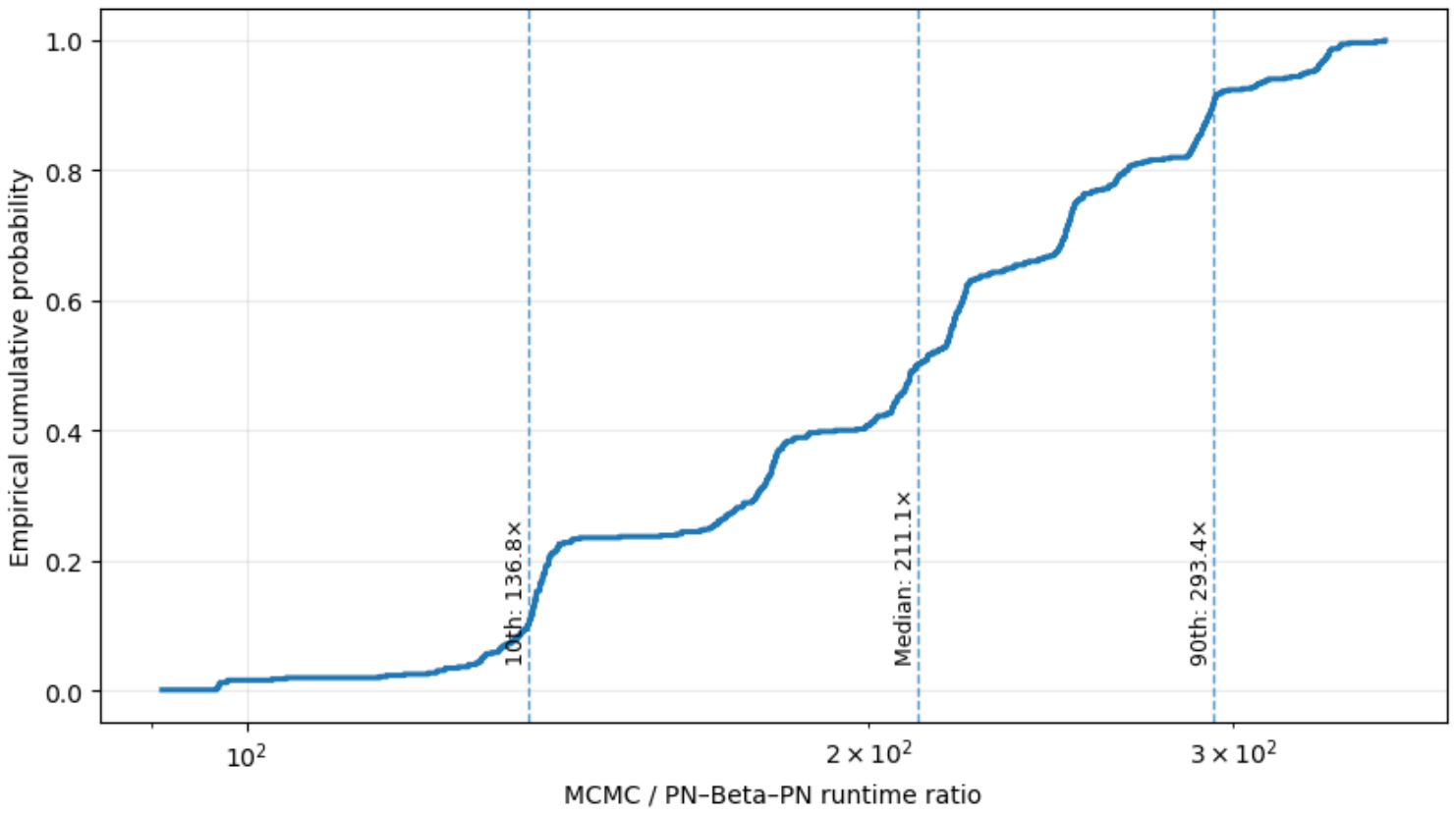

Figure A. 3. Empirical cumulative distribution of the converged-MCMC-to-PN–Beta–PN per-update runtime ratio over the full numerical experiment.

Overall, the PN–Beta–PN procedure closely reproduces the converged quadrature posterior over most of the tested domain while avoiding repeated posterior integration and sampling. Its largest discrepancies are concentrated in nearly deterministic boundary regimes, where posterior exceedance probabilities are already close to zero or one. In the intermediate-probability region most relevant to uncertainty-sensitive fragility updating, the approximation provides a computationally efficient deterministic alternative to repeated quadrature while being a substantially faster alternative.

This computational advantage becomes especially important when local updates must be repeated across large building inventories, multiple damage thresholds, heterogeneous evidence streams, and sequential data-assimilation batches.

## References

[1] Cornell CA. Bounds on the reliability of structural systems. Journal of the Structural Division. 1967;93:171-200.
[2] Ang AH. Structural risk analysis and reliability-based design. Journal of the Structural Division. 1973;99:1891-910.
[3] Ang AH, Cornell CA. Reliability bases of structural safety and design. Journal of the structural division. 1974;100:1755-69.
[4] Ellingwood B. Development of a probability based load criterion for American National Standard A58: Building code requirements for minimum design loads in buildings and other structures: US Department of Commerce, National Bureau of Standards; 1980.
[5] Kennedy R, Short SA. Basis for seismic provisions of DOE-STD-1020. Lawrence Livermore National Lab. (LLNL), Livermore, CA (United States …; 1994.
[6] Kircher CA, Nassar AA, Kustu O, Holmes WT. Development of building damage functions for earthquake loss estimation. Earthquake spectra. 1997a;13:663-82.

[7] Kircher CA, Reitherman RK, Whitman RV, Arnold C. Estimation of earthquake losses to buildings. Earthquake spectra. 1997b;13:703-20.
[8] FEMA. HAZUS technical manual. National Institute of Building for the Federal Emergency Management Agency, Washington (DC). 1997.
[9] Koliou M, van de Lindt JW, McAllister TP, Ellingwood BR, Dillard M, Cutler H. State of the research in community resilience: Progress and challenges. Sustainable and resilient infrastructure. 2020;5:131-51.
[10] Burton H, Rad AR, Yi Z, Gutierrez D, Ojuri K. Seismic collapse performance of Los Angeles soft, weak, and open-front wall line woodframe structures retrofitted using different procedures. Bulletin of earthquake engineering. 2019;17:2059-91.
[11] Choe D-E, Gardoni P, Rosowsky D, Haukaas T. Probabilistic capacity models and seismic fragility estimates for RC columns subject to corrosion. Reliability Engineering & System Safety. 2008;93:383-93.
[12] Ramamoorthy SK, Gardoni P, Bracci JM. Probabilistic demand models and fragility curves for reinforced concrete frames. Journal of structural Engineering. 2006;132:1563-72.
[13] Gardoni P, Der Kiureghian A, Mosalam KM. Probabilistic capacity models and fragility estimates for reinforced concrete columns based on experimental observations. Journal of Engineering Mechanics. 2002;128:1024-38.
[14] Qiu C-X, Zhu S. Performance-based seismic design of self-centering steel frames with SMA-based braces. Engineering Structures. 2017;130:67-82.
[15] Li Y, Song R, Van De Lindt JW. Collapse fragility of steel structures subjected to earthquake mainshock-aftershock sequences. Journal of Structural Engineering. 2014;140:04014095.
[16] Cornell CA, Jalayer F, Hamburger RO, Foutch DA. Probabilistic basis for 2000 SAC federal emergency management agency steel moment frame guidelines. Journal of structural engineering. 2002;128:526-33.
[17] Rosowsky DV, Ellingwood BR. Performance-based engineering of wood frame housing: Fragility analysis methodology. Journal of Structural Engineering. 2002;128:32-8.
[18] Ellingwood BR, Rosowsky DV, Li Y, Kim JH. Fragility assessment of light-frame wood construction subjected to wind and earthquake hazards. Journal of structural engineering. 2004;130:1921-30.
[19] Filiatrault A, Sullivan T. Performance-based seismic design of nonstructural building components: The next frontier of earthquake engineering. Earthquake Engineering and Engineering Vibration. 2014;13:17-46.
[20] Vamvatsikos D, Cornell CA. Incremental dynamic analysis. Earthquake engineering & structural dynamics. 2002;31:491-514.
[21] Abdelhady AU, Spence SM, McCormick J. Risk and fragility assessment of residential wooden buildings subject to hurricane winds. Structural Safety. 2022;94:102137.
[22] Li Y, Ellingwood BR. Hurricane damage to residential construction in the US: Importance of uncertainty modeling in risk assessment. Engineering structures. 2006;28:1009-18.
[23] Memari M, Attary N, Masoomi H, Mahmoud H, van de Lindt JW, Pilkington SF et al. Minimal building fragility portfolio for damage assessment of communities subjected to tornadoes. Journal of Structural Engineering. 2018;144:04018072.
[24] Nofal OM, van de Lindt JW. Fragility-based flood risk modeling to quantify the effect of policy change on losses at the community level. Civ Eng Res J. 2021;11:555822.
[25] van de Lindt JW, Taggart M. Fragility analysis methodology for performance-based analysis of wood-frame buildings for flood. Natural Hazards Review. 2009;10:113-23.
[26] Nofal OM, van de Lindt JW, Do TQ. Multi-variate and single-variable flood fragility and loss approaches for buildings. Reliability Engineering & System Safety. 2020;202:106971.
[27] Tomiczek T, Kennedy A, Rogers S. Collapse limit state fragilities of wood-framed residences from storm surge and waves during Hurricane Ike. Journal of Waterway, Port, Coastal, and Ocean Engineering. 2014;140:43-55.
[28] Do TQ, van de Lindt JW, Cox DT. Hurricane surge-wave building fragility methodology for use in damage, loss, and resilience analysis. Journal of structural engineering. 2020;146:04019177.
[29] Masoomi H, van de Lindt JW. Tornado fragility and risk assessment of an archetype masonry school building. Engineering Structures. 2016;128:26-43.
[30] Koliou M, Masoomi H, van de Lindt JW. Performance assessment of tilt-up big-box buildings subjected to extreme hazards: Tornadoes and earthquakes. Journal of Performance of Constructed Facilities. 2017;31:04017060.
[31] Masoomi H, Ameri MR, van de Lindt JW. Wind performance enhancement strategies for residential wood-frame buildings. Journal of Performance of Constructed Facilities. 2018;32:04018024.
[32] Padgett JE, DesRoches R. Methodology for the development of analytical fragility curves for retrofitted bridges. Earthquake Engineering & Structural Dynamics. 2008;37:1157-74.
[33] Nie Y, Li J, Liu G, Zhou P. Cascading failure-based reliability assessment for post-seismic performance of highway bridge network. Reliability Engineering & System Safety. 2023;238:109457.

[34] Ma X, Xiong W, Zhou D, Cai C. Fragility analysis of multiple bridges under scour conditions enabled by knowledge transferred from single bridge. Reliability Engineering & System Safety. 2025:111837.
[35] Braik AM, Salman AM, Li Y. Risk-based reliability and cost analysis of utility poles subjected to tornado hazard. Journal of Aerospace Engineering. 2019;32:04019040.
[36] Li Y, Salman AM, Braik A, Bjarnadóttir S, Salarieh B. Risk-Based Management of Electric Power Distribution Systems Subjected to Hurricane and Tornado Hazards. Engineering for Extremes: Decision-Making in an Uncertain World: Springer; 2021. p. 143-66.
[37] Salman AM, Li Y, Bastidas-Arteaga E. Maintenance optimization for power distribution systems subjected to hurricane hazard, timber decay and climate change. Reliability Engineering & System Safety. 2017;168:136-49.
[38] Zhang J, Bagtzoglou Y, Zhu J, Li B, Zhang W. Fragility-based system performance assessment of critical power infrastructure. Reliability Engineering & System Safety. 2023;232:109065.
[39] Farahmandfar Z, Piratla KR, Andrus RD. Resilience evaluation of water supply networks against seismic hazards. Journal of Pipeline Systems Engineering and Practice. 2017;8:04016014.
[40] Zhang Y, Zhong Z, Li J, Hou B, Xu C. Experimental investigation and seismic performance assessment of cast iron pipelines considering the mechanical property uncertainties of cement-caulked joints. Reliability Engineering & System Safety. 2025:111625.
[41] FEMA. HAZUS-MHMR Methodology, Multi-hazard Loss Estimation. Technical Manual, NIST, Washington, DC. 2003.
[42] FEMA. Seismic Performance Assessment of Buildings, Volume 1 – Methodology. FEMA P-58-1, prepared by the Applied Technology Council (ATC) for the Federal Emergency Management Agency. 2012.
[43] Braik AM, Gupta HS, Koliou M, González AD. Multi-hazard probabilistic risk assessment and equitable multi-objective optimization of building retrofit strategies in hurricane-vulnerable communities. Computer-Aided Civil and Infrastructure Engineering. 2025;40:2074–97.
[44] Aghababaei M, Koliou M, Paal SG. Performance assessment of building infrastructure impacted by the 2017 Hurricane Harvey in the Port Aransas region. Journal of Performance of Constructed Facilities. 2018;32:04018069.
[45] Van de Lindt JW, Peacock WG, Mitrani-Reiser J, Rosenheim N, Deniz D, Dillard M et al. Community resilience-focused technical investigation of the 2016 Lumberton, North Carolina, flood: An interdisciplinary approach. Natural Hazards Review. 2020;21:04020029.
[46] Aghababaei M, Okamoto C, Koliou M. Examining the accuracy and validity of loss estimations using the PBEE methodology for wood residential buildings through integrated experimental findings and expert panel solicitation. Sustainable and Resilient Infrastructure. 2022;7:638-54.
[47] van de Lindt JW, Wang WL, Johnston B, Crawford PS, Yan G, Dao T et al. Social Susceptibility–Driven Longitudinal Tornado Reconnaissance Methodology: 2021 Midwest Quad-State Tornado Outbreak. ASCE OPEN: Multidisciplinary Journal of Civil Engineering. 2025;3:04025006.
[48] Merhi A, Brown-Giammanco TM, Roueche DB, Wood RL, Murphy TA, Standohar-Alfano CD et al. Damage Assessment of the 2020 Monroe, Louisiana, Tornado Using Different Remote-Sensing Methods and Enhanced Fujita Scales. Natural Hazards Review. 2026;27:04025062.
[49] Matin SS, Pradhan B. Challenges and limitations of earthquake-induced building damage mapping techniques using remote sensing images-A systematic review. Geocarto International. 2022;37:6186-212.
[50] Kaur N, Lee CC, Mostafavi A, Mahdavi-Amiri A. Large-scale building damage assessment using a novel hierarchical transformer architecture on satellite images. Computer-Aided Civil and Infrastructure Engineering. 2023.
[51] Kaur S, Gupta S, Singh S, Koundal D, Zaguia A. Convolutional neural network based hurricane damage detection using satellite images. Soft Computing. 2022;26:7831-45.
[52] Khajwal AB, Cheng CS, Noshadravan A. Post-disaster damage classification based on deep multi-view image fusion. Computer-Aided Civil and Infrastructure Engineering. 2023;38:528-44.
[53] Braik AM, Koliou M. Automated building damage assessment and large-scale mapping by integrating satellite imagery, GIS, and deep learning. Computer-Aided Civil and Infrastructure Engineering. 2024a;39, 2389–2404.
[54] Fan C, Jiang Y, Mostafavi A. Social sensing in disaster city digital twin: Integrated textual–visual–geo framework for situational awareness during built environment disruptions. Journal of Management in Engineering. 2020;36:04020002.
[55] Ford DN, Wolf CM. Smart cities with digital twin systems for disaster management. Journal of management in engineering. 2020;36:04020027.
[56] Ghosh S, Huyck CK, Greene M, Gill SP, Bevington J, Svekla W et al. Crowdsourcing for rapid damage assessment: The global earth observation catastrophe assessment network (GEO-CAN). Earthquake Spectra. 2011;27:179-98.

[57] Khajwal AB, Noshadravan A. An uncertainty-aware framework for reliable disaster damage assessment via crowdsourcing. International Journal of Disaster Risk Reduction. 2021;55:102110.
[58] Cheng C-S, Behzadan AH, Noshadravan A. Bayesian inference for uncertainty-aware post-disaster damage assessment using artificial intelligence. Computing in Civil Engineering 20212022. p. 156-63.
[59] Behrooz H, Ilbeigi M, Reisi-Gahrooei M. Adaptive Data Collection for Post-Disaster Rapid Damage Assessment: A Multi-Objective Optimization Approach. Available at SSRN 5274318.
[60] Huang S, Sun C, Gong J, Pompili D. Reinforcement learning–based task allocation and path-finding in multi-robot systems under environment uncertainty. Computer-Aided Civil and Infrastructure Engineering. 2025.
[61] Braik AM, Koliou M. Post-Tornado Automated Building Damage Evaluation and Recovery Prediction by Integrating Remote Sensing, Deep Learning, and Restoration Models. Sustainable Cities and Society. 2025b;106286.
[62] Braik AM, Han X, Koliou M. A Framework for Resilience Analysis and Equitable Recovery in Tornado-impacted Communities Using Agent-based Modeling and Computer Vision-based Damage Assessment. International Journal of Disaster Risk Reduction. 2025:105427.
[63] De Risi R, Goda K, Mori N, Yasuda T. Bayesian tsunami fragility modeling considering input data uncertainty. Stochastic environmental research and risk assessment. 2017;31:1253-69.
[64] Koutsourelakis PS. Assessing structural vulnerability against earthquakes using multi-dimensional fragility surfaces: A Bayesian framework. Probabilistic Engineering Mechanics. 2010;25:49-60.
[65] Braik AM, Koliou M. A novel digital twin framework of electric power infrastructure systems subjected to hurricanes. International Journal of Disaster Risk Reduction. 2023;97:104020.
[66] Braik AM, Koliou M. A digital twin framework for efficient electric power restoration and resilient recovery in the aftermath of hurricanes considering the interdependencies with road network and essential facilities. Resilient Cities and Structures. 2024c;3:79-91.
[67] Braik AM, Koliou M. Spatially Correlated Multi-State Fragility via a Warped Gaussian Process. Structural Safety. 2026;121:102706.
[68] Johnson NL, Kotz S, Balakrishnan N. Continuous univariate distributions, volume 2: John wiley & sons; 1995.
[69] MacKay DJ. Information theory, inference and learning algorithms: Cambridge university press; 2003.
[70] Cover TM. Elements of information theory: John Wiley & Sons; 1999.
[71] van de Lindt JW, Peacock WG, Mitrani-Reiser J, Rosenheim N, Deniz D, Dillard M et al. The Lumberton, North Carolina Flood of 2016: A community resilience focused technical investigation. 2018.
[72] Ren P, Chen X, Sun L, Sun H. Incremental Bayesian matrix/tensor learning for structural monitoring data imputation and response forecasting. Mechanical Systems and Signal Processing. 2021;158:107734.
[73] Gelman A, Carlin JB, Stern HS, Rubin DB. Bayesian data analysis: Chapman and Hall/CRC; 1995.
[74] Gelfand AE, Diggle P, Guttorp P, Fuentes M. Handbook of spatial statistics: CRC press; 2010.
[75] Williams CK, Rasmussen CE. Gaussian processes for machine learning: MIT press Cambridge, MA; 2006.
[76] Titsias M. Variational learning of inducing variables in sparse Gaussian processes. Artificial intelligence and statistics: PMLR; 2009. p. 567-74.
[77] Hensman J, Fusi N, Lawrence ND. Gaussian processes for big data. arXiv preprint arXiv:13096835. 2013.
[78] IN-CORE. Joplin Testbed. 2025.
[79] Kuligowski ED, Lombardo FT, Phan L, Levitan ML, Jorgensen DP. Final report, National Institute of Standards and Technology (NIST) technical investigation of the May 22, 2011, tornado in Joplin, Missouri. 2014.
[80] NOAA. National Oceanic and Atmospheric Administration. Joplin Tornado Images. 2011.
[81] Standohar-Alfano CD, van de Lindt JW. Empirically based probabilistic tornado hazard analysis of the United States using 1973–2011 data. Natural Hazards Review. 2015;16:04014013.
[82] WSEC. A recommendation for an Enhanced Fujita Scale. Wind Science and Engineering Center (TTU WiSE), Texas Tech University, Lubbock, Texas. 2006.
[83] FEMA. Hazus Earthquake Model Technical Manual-Hazus 6.1. Federal Emergency Management Agency: Washington, DC, USA. 2024.
[84] Gupta R, Hosfelt R, Sajeev S, Patel N, Goodman B, Doshi J et al. xbd: A dataset for assessing building damage from satellite imagery. arXiv preprint arXiv:191109296. 2019.
[85] Braik AM, Salman AM, Li Y. Reliability-based assessment and cost analysis of power distribution systems at risk of Tornado hazard. ASCE-ASME Journal of Risk and Uncertainty in Engineering Systems, Part A: Civil Engineering. 2020;6:04020014.
[86] Gardner J, Pleiss G, Weinberger KQ, Bindel D, Wilson AG. Gpytorch: Blackbox matrix-matrix gaussian process inference with gpu acceleration. Advances in neural information processing systems. 2018;31.